\documentclass[smallcondensed]{svjour3}     
\smartqed  
\usepackage{graphicx}
\usepackage{amssymb}
\usepackage[figuresright]{rotating}
\usepackage{natbib}
\usepackage{hyperref}
\hypersetup{colorlinks=true,
            citecolor=blue,
            linkcolor=blue}

\def\nul#1{}
\def \vv#1{\mathbf{#1}}
\def \be {\begin{equation}}
\def \ee {\end{equation}}
\def \k  {s}
\def \K  {k}
\def \p  {0}
\def \s  {1}
\def \e {2}
\def \a  {2}
\def \b  {1}

\def \m  {m_{01}}
\def \M  {m_\e}
\def \w {\omega}
\def \om {\Omega}
\def \ij {i}
\def \ji {j}
\def \jk {j}

\def \eq {{}}
\def \de {\theta}
\def \ve {\varepsilon}

\def\ui{u_\p}
\def\vi{v_\p}
\def\uo{u_\e}
\def\vo{v_\e}
\def\A {{\cal A}}
\def\B {{\cal B}}
\def\C {{\cal C}}
\def\D {{\cal D}}
\def\E {{\cal E}}
\def\F {{\cal F}}
\def\G {\mathbb{G}}
\def\Z {\mathbb{Z}}

\def\figpath{}

\def\norm#1{\left\Vert#1\right\Vert}

\def\Frac#1#2{{{\displaystyle\strut#1}\over{\displaystyle\strut#2}}}

\def\crm{\cr\noalign{\medskip}}
\def\trans#1{{}{#1}^t}
\def\moy#1{\left\langle{#1}\right\rangle}
\def\d{{\rm d}}

\def \llabel#1{\label{#1}}

\def\EQM#1{\vcenter{\normalbaselines\m@th
    \ialign{${\displaystyle ##}$\hfil&&\ ${\displaystyle ##}$\hfil\crcr
    \mathstrut\crcr\noalign{\kern-\baselineskip}
    \noalign{\smallskip}
    #1\crcr\mathstrut\crcr\noalign{\kern-\baselineskip}}}}

\def\bfx#1{{#1}}

\begin{document}
\title{Secular and tidal evolution of circumbinary systems.} 


\author{Alexandre C. M. Correia         \and
        Gwena\"el Bou\'e               \and 
        Jacques Laskar                 
}


\institute{A.C.M. Correia \at
              CIDMA, Departamento de F\'isica, Universidade de Aveiro, 
              Campus de Santiago, 3810-193 Aveiro, Portugal 
             \email{correia@ua.pt}           
           \and
	   G. Bou\'e 
           \and
           J. Laskar  \at
           ASD, IMCCE-CNRS UMR8028, Obs. Paris,
           77 Av. Denfert-Rochereau, 75014 Paris, France 
}

\date{\today}

\maketitle

\begin{abstract}
We investigate the secular dynamics of three-body circumbinary systems under the effect of tides.
We use the octupolar non-restricted approximation for the orbital interactions, general relativity corrections, 
the quadrupolar approximation for the spins, and the viscous linear model for tides. 
We derive the averaged equations of motion in a simplified vectorial formalism, which is suitable to model the long-term evolution of a wide variety of circumbinary systems in very eccentric and inclined orbits.
In particular, this vectorial approach can be used to derive constraints for tidal migration, capture in Cassini states, and stellar spin-orbit misalignment.
We show that circumbinary planets with initial arbitrary orbital inclination can become coplanar through a secular resonance between the precession of the orbit and the precession of the spin of one of the stars.
We also show that circumbinary systems for which the pericenter of the inner orbit is initially in libration present chaotic motion for the spins and for the eccentricity of the outer orbit.
Because our model is valid for the non-restricted problem, it can also be applied to any three-body hierarchical system such as star-planet-satellite systems and triple stellar systems.

\keywords{Extended Body \and Dissipative Forces \and
Planetary Systems \and Rotation}  
\end{abstract}



\section{Introduction}

Circumbinary bodies are objects that orbit around a more massive binary system.
In the solar system, the small satellites of the Pluto-Charon system are the best example \citep[e.g.][]{Brozovic_etal_2015}.
Planets orbiting two stars, often called circumbinary planets, have also been reported, at present we know about 20 of them\footnote{http://exoplanet.eu/}.
These kind of systems are particularly interesting from a dynamical point of view, as they can be stable for very eccentric and inclined orbits and thus present uncommon behaviors.
Moreover, many circumbinary systems are observed within 1~AU, therefore they can undergo tidal dissipation, which slowly modifies the spins and the orbits, in particular for the inner binary pair \citep[e.g.][]{MacDonald_1964}.
As a consequence, the final configuration of these systems can be totally different from the initial one \citep[e.g.][]{Correia_etal_2011}.

Half of the already known circumbinary planets were detected using the {\it Kepler Spacecraft}\footnote{http://kepler.nasa.gov/} data, which is based on the transits observational technique \citep[e.g.][]{Doyle_etal_2011, Welsh_etal_2012, Orosz_etal_2012a, Orosz_etal_2012b, Kostov_etal_2016}. 
These systems are thus almost coplanar, although this is not necessarily the standard configuration for circumbinary planets \citep[e.g.][]{Martin_Triaud_2014}.
\bfx{Indeed, misaligned circumbinary discs were already observed \citep[e.g.][]{Kennedy_etal_2012a, Plavchan_etal_2013}. }
Moreover, for systems detected with different observational techniques, the mutual inclination is usually not constrained, but it is compatible with no coplanar orbits \citep[e.g.][]{Ford_etal_2000p, Couetdic_etal_2010}.

The origin and evolution of circumbinary systems can be analyzed with direct numerical integrations of the full equations of motion \citep[e.g.][]{Verrier_Evans_2009, Martin_Triaud_2014}, but a theoretical understanding of the dynamics often requires the development and application of theories with valid analytic approximations. 
Additionally, tidal effects usually act over very long time-scales and therefore approximate theories also allow to speed-up the numerical simulations and to explore the parameter space much more rapidly.

Stability studies have shown that for semi-major axis ratios $a_\b / a_\a > 1/2$ 
there is a large overlap of mean motion resonances and circumbinary orbits are most likely unstable \citep[e.g.][]{Doolin_Blundell_2011, Bosanac_etal_2015, Correia_etal_2015}.
For smaller ratios we can use secular perturbation theories based on series development in the ratio of the semi-major axis $a_\b / a_\a $.
The development to the second order, called the quadrupole approximation, was used by \citet{Lidov_1962} and \citet{Kozai_1962} for the restricted inner problem (the outer orbit is unperturbed).
\citet{Farago_Laskar_2010} derived a simple model for the non-restricted quadrupolar problem of three masses and \citet{Correia_etal_2011} added the effect from tides.
However, the quadrupole approximation is insufficient for studying nearly coplanar systems, because it fails to reproduce the eccentricity oscillations.
For that purpose we need to extend the series development in $a_\b / a_\a $ to the third order, that is, to the octopole order \citep[e.g.][]{Marchal_1990, Ford_etal_2000, Laskar_Boue_2010}.

In this paper we intend to get deeper into the study of circumbinary three-body
systems, where all bodies undergo tidal interactions.
We do not make any restrictions on the masses of these bodies, and use the octupolar approximation for the orbital gravitational interactions with general relativity corrections.
It has been shown that the tidal evolution of the orbits cannot be dissociated from the spin evolution \citep[e.g.][]{Correia_etal_2012}.
Therefore, we also consider the full effect on the spins of all bodies (up to the quadrupolar approximation), including the rotational flattening of their figures.
This allows us to correctly describe the precession of the spin axis and subsequent capture in Cassini states.
We adopt a viscous linear model for tides \citep{Singer_1968,Mignard_1979}, as it provides simple expressions for the tidal torques for any eccentricity value.
For gaseous planets and stars, this model has also shown to be a very good approximation \citep[e.g.][]{Ferraz-Mello_2013, Correia_etal_2014}.
Since we are interested in the secular behavior, we average the motion equations over the mean anomalies of the orbits and express them using the vectorial methods \citep[e.g.][]{Boue_Laskar_2006, Boue_Laskar_2009, Correia_2009, Farago_Laskar_2010, Boue_Fabrycky_2014b}. 

In Section\,\ref{secmodel} we derive the averaged equations of motion that we use to evolve circumbinary systems by tidal effect. 
In Section\,\ref{secevol} we obtain the secular evolution of the spin and orbital quantities in terms of reference angles and elliptical elements, that are useful and more intuitive to understand the outcomes of the numerical simulations.
In Section\,\ref{tidalevol} we include the contribution of tidal effects and perform some numerical simulations to illustrate some unexpected behaviors for circumbinary systems.
In Section\,\ref{otherappli} we explain how our model can be extended to any three-body hierarchical system.
Finally, last section is devoted to the conclusions.

\section{Model}
\llabel{secmodel}

We consider here a system composed of an inner pair of bodies with stellar masses $m_\p$ and
$m_\s$, together with an external planetary companion with mass $\M$:
\begin{equation}
m_\p \ge m_\b \gg \M \ . \llabel{160608b}
\end{equation}
All bodies are considered oblate ellipsoids with gravity field
coefficients given by $J_{2,\ij}$, rotating about the axis of maximal
inertia along the directions $\vv{\hat \k}_\ij$ (gyroscopic approximation), with rotation
rates $\om_\ij$, such that
\citep[e.g.][]{Lambeck_1988}
\begin{equation}
J_{2,\ij} = k_{2,\ij} \frac{\om_\ij^2 R_\ij^3}{3 G m_\ij} \ ,
\llabel{101220a}
\end{equation}
where $G$ is the gravitational constant, $R_\ij$ is the radius of each
body, and $ k_{2,\ij} $ is the second Love number for potential.
The index $i = 0, 1, 2$ pertains to the body with mass $m_i$.

In order to obtain the equations of motion we use Jacobi canonical
coordinates, which are the distance between the two innermost bodies,
$\vv{r}_\b$, and the distance from the external body to the inner's orbit
center of mass, $ \vv{r}_\a $ (see Fig.~\ref{fig1}). 
We additionally assume that $|\vv{r}_\b|  \ll  |\vv{r}_\a|$ and adopt the
octupolar three-body problem approximation. 
In the following, for any vector $\vv{u}$, $\vv{\hat u} = \vv{u} / \norm{\vv{u}}$ is the unit vector. 

\begin{figure}
\begin{center}
\includegraphics[width=\textwidth]{\figpath 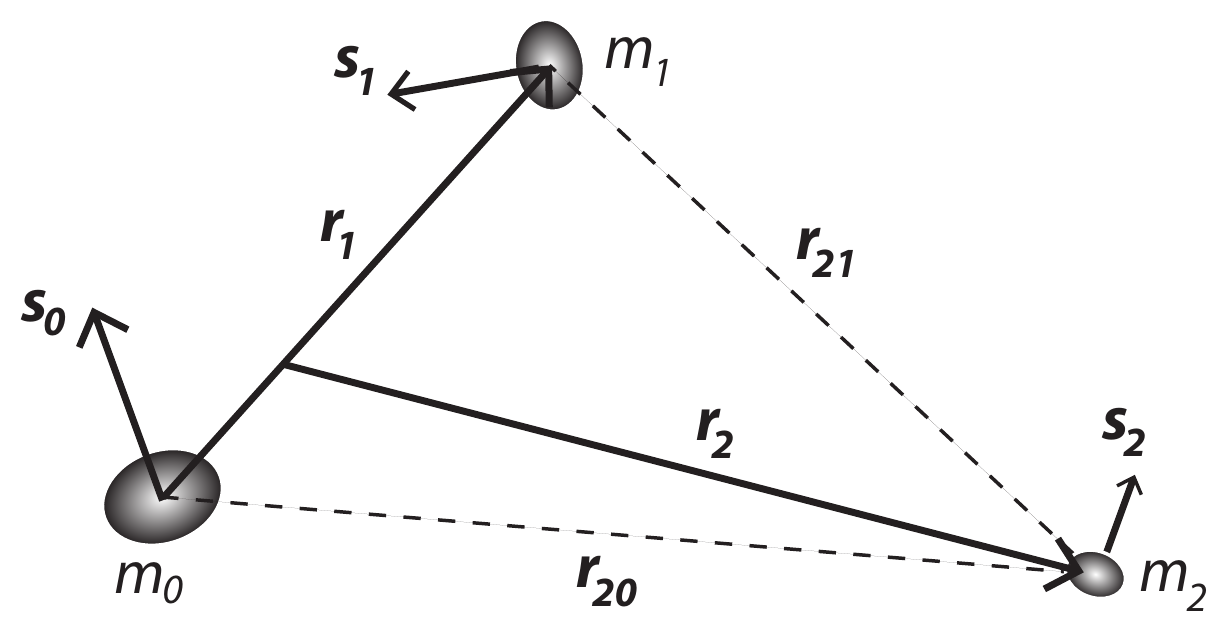} 
 \caption{Jacobi coordinates, where $\vv{r}_\b $ is the position of $m_\s$ relative to $m_\p$ (inner orbit), and $ \vv{r}_\a $ the position of $\M$ relative to the center of mass of $m_\p$ and $m_\s$ (outer orbit). All bodies are considered oblate ellipsoids, where $\vv{\hat \k}_\ij$ is the spin axis. \llabel{fig1}  }
\end{center}
\end{figure}

\subsection{Orbital motion}
\llabel{orbmotion}

The potential energy $U$ of a hierarchical three-body system of punctual masses is given in Jacobi coordinates by 
\citep[e.g.][]{Smart_1953}: 
\be
U =  
- G \frac{m_\p m_\s}{r_\b} 
- G \frac{\M \m}{r_\a} + U' \ ;
\ee
with
\be
U'=- G \frac{\M \beta_\b}{r_\a} \left(\frac{r_\b}{r_\a}\right)^2 \!\! P_2 (\vv{\hat
r}_\a \cdot \vv{\hat r}_\b) 
- G \frac{\M \beta_\b}{r_\a} \frac{m_\p - m_\s}{m_\p + m_\s}
\left(\frac{r_\b}{r_\a}\right)^3 \!\! P_3 (\vv{\hat r}_\a \cdot \vv{\hat r}_\b) 
\ , \llabel{090514a}
\ee
where $P_2(x) = (3x^2-1)/2$ and $P_3(x) = (5x^3-3x)/2$ are the Legendre polynomial of degree two and three, respectively, and terms in $(r_\b/r_\a)^4$ have been neglected.
We also have $\m = (m_\p + m_\s) $,
$ \beta_\b = m_\p m_\s / \m $,  $ \beta_\a = \M \m / (\M + \m) $, $\mu_\b=G \m$,
and $\mu_\a=G(\M + \m)$.
The evolution of the orbits can be tracked by the orbital angular momenta, 
\be
\vv{G}_\ij =  G_\ij\, \vv{\hat \K}_\ij \ ,   \hbox{with} \quad G_\ij=\beta_\ij \sqrt{\mu_\ij a_\ij (1-e_\ij^2)}
\llabel{130116c}
\ee
where $\vv{\hat \K}_\ij$ is the unit vector $\vv{\hat G}_\ij$, $a_\ij$ is the semi-major axis, and $e_\ij$ is the eccentricity.  
The mean motion is defined as $n_\ij = \sqrt{\mu_\ij / a_\ij^3 }$. 
The Laplace-Runge-Lenz vector $\vv{e}_\ij$ points along the major axis in the direction of periapsis with  magnitude $e_\b$ and is expressed as  
\be
\vv{e}_\ij = \Frac{\vv{\dot r}_\ij \times \vv{G}_\ij}{\beta_\ij  \mu_\ij} -
\Frac{\vv{r}_\ij}{r_\ij} \llabel{100119a} \ .
\ee
The contributions to the orbits are easily computed from the potential as
\be
\vv{\dot G}_\ij = \vv{r}_\ij \times \vv{F}_\ij \ , \quad
\llabel{090514d}
\ee
and
\be
\dot \vv{e}_\ij = \Frac{1}{\beta_\ij \mu_\ij} \left( \vv{F}_\ij
\times \Frac{\vv{G}_\ij}{\beta_\ij} + \vv{\dot r}_\ij \times \vv{\dot G}_\ij \right)
\llabel{100119b}  \ ,
\ee
where $ \vv{F}_\ij = - \vv{\nabla}_{\!\vv{r}_\ij} U' $. 

Because we are only interested in the secular evolution of the system, we
further average the equations of motion over the mean
anomalies of both orbits (see appendix~\ref{apenA}). 
The resulting equations are:

\paragraph{Quadrupole:}
\begin{eqnarray}
\vv{\dot G}_\b = - \gamma_2 \Big[ (1-e_\b^2) \cos I \, \vv{\hat \K}_\a \times \vv{\hat
\K}_\b - 5 (\vv{e}_\b \cdot \vv{\hat \K}_\a) \, \vv{\hat \K}_\a \times \vv{e}_\b \Big]
 \ , \llabel{090514z2}
\end{eqnarray}
\begin{eqnarray}
\vv{\dot G}_\a = - \vv{\dot G}_\b 
 \ , \llabel{090514z1}
\end{eqnarray}
\begin{eqnarray}
\vv{\dot e}_\b &=& - \Frac{\gamma_2 (1-e_\b^2)}{\norm{\vv{G}_\b}} \Big[ \cos I \,
\vv{\hat \K}_\a \times \vv{e}_\b - 2 \, \vv{\hat \K}_\b \times \vv{e}_\b - 5
(\vv{e}_\b \cdot \vv{\hat \K}_\a) \, \vv{\hat \K}_\a \times \vv{\hat \K}_\b \Big]
 \ , \llabel{090514z3}
\end{eqnarray}
\begin{eqnarray}
\vv{\dot e}_\a &=& - \Frac{\gamma_2}{\norm{\vv{G}_\a}} \Big[
(1-e_\b^2)  \cos I \, \vv{\hat \K}_\b \times \vv{e}_\a 
- 5 (\vv{e}_\b \cdot \vv{\hat \K}_\a) \, \vv{e}_\b \times \vv{e}_\a
\phantom{\frac{1}{2}} \Big. 
\crm &&
\Big. + \frac{1}{2} \left( 1 - 6 e_\b^2 - 5 (1-e_\b^2) \cos^2 I + 25 (\vv{e}_\b \cdot
\vv{\hat \K}_\a)^2 \right) \, \vv{\hat \K}_\a \times \vv{e}_\a \Big]
 \ , \llabel{110506a}
\end{eqnarray}
with
\be
\cos I   = \vv{\hat \K}_\b \cdot \vv{\hat \K}_\a \ , \llabel{130117a}
\ee
and
\be 
\gamma_2 = \frac{3 G \M \beta_\b a_\b^2}{4 a_\a^3 (1-e_\a^2)^{3/2}} \ .
\llabel{090514i}
\ee

\paragraph{Octupole:}

\begin{eqnarray}
\vv{\dot G}_\b = \gamma_3 \Big[
(\B\, \vv{e}_\a + \C\, \vv{\hat \K}_\a) \times \vv{e}_\b
+ (\D\, \vv{e}_\a + \E\, \vv{\hat \K}_\a) \times \vv{\hat \K}_\b \Big]
 \ , \llabel{120117b}
\end{eqnarray}
\begin{eqnarray}
\vv{\dot G}_\a = - \vv{\dot G}_\b
 \ , \llabel{120117c}
\end{eqnarray}
\begin{eqnarray}
\vv{\dot e}_\b &=&\frac{\gamma_3}{\norm{\vec G_1}} \Big[
(1-e_\b^2) (\A\, \vv{e}_\b + \B\, \vv{e}_\a + \C\, \vv{\hat \K}_\a) 
\times \vv{\hat \K}_\b + (\D\, \vv{e}_\a + \E\, \vv{\hat \K}_\a) \times \vv{e}_\b
\Big] \ , \llabel{120117d}
\end{eqnarray}
\begin{eqnarray}
\vv{\dot e}_\a &=& \frac{\gamma_3}{\norm{\vec G_2}} \Big[
\left( \F + \C (\vv{e}_\b \cdot \vv{\hat \K}_\a) + \E \cos I \right) \, \vv{e}_\a  \times \vv{\hat \K}_\a 
\crm & &
+ (1-e_2^2) (\B\, \vv{e}_\b + \D\, \vv{\hat \K}_\b) \times \vv{\hat \K}_\a 
+ (\C\, \vv{e}_\b + \E\, \vv{\hat \K}_\b) \times \vv{e}_\a
\Big] \ , \llabel{120117e}
\end{eqnarray}

with

\begin{eqnarray}
\A &=& 16 (\vv{e}_\b \cdot \vv{e}_\a) \ , \nonumber \\ 
\B &=& - \Big[ 1 - 5 (1-e_1^2) \cos^2 I
    + 35 (\vv{e}_\b \cdot \vv{\hat \K}_\a)^2 - 8 e_1^2 \Big] \ ,\nonumber  \\ 
\C &=& 10 (1-e_1^2)  (\vv{\hat \K}_\b \cdot \vv{e}_\a) \cos I
    - 70 (\vv{e}_\b \cdot \vv{\hat \K}_\a) (\vv{e}_\b \cdot \vv{e}_\a) \ , \nonumber \\ 
\D &=& 10 (1-e_1^2) (\vv{e}_\b \cdot \vv{\hat \K}_\a) \cos I \ ,  \\ 
\E &=& 10 (1-e_1^2) \Big[   (\vv{e}_\b \cdot  \vv{e}_\a) \cos I
  + (\vv{e}_\b \cdot \vv{\hat \K}_\a) (\vv{\hat \K}_\b \cdot \vv{e}_\a) \Big] \ , \nonumber \\ 
\F &=& 5 \Big[
\B (\vv{e}_\b \cdot \vv{e}_\a ) + \D  (\vv{\hat \K}_\b \cdot \vv{e}_\a)
 \Big] \nonumber
\ , \llabel{120117z}
\end{eqnarray}
and
\be 
\gamma_3 = \frac{15}{64} \frac{G m_2 \beta_\b a_\b^3}{a_\a^4
(1-e_\a^2)^{5/2}} \frac{m_\p-m_\s}{m_\p+m_\s} \ .
\llabel{120117a}
\ee

\subsection{General relativity correction}

We can add to Newton's equations  the dominant contributions from
general relativistic effects. 
These effects are mainly felt by eccentric orbits on close encounters between the central stars, and contribute to the gravitational force with a small correction
\citep[e.g.][]{Kidder_1995}
\be
\vv{F}_\mathrm{gr} = - \frac{\beta_\b \mu_\b}{c^2 r_\b^2} \left[ \left( (1+3 \eta) \vv{\dot r}_\b^2 - 2 (2+\eta)\frac{\mu_\b}{r_\b} -\frac32 \eta \, \dot r_\b^2 \right) \hat \vv{r}_\b 
-2 (2-\eta) \dot r_\b \vv{\dot r}_\b \right] \ , \llabel{160503a}
\ee
where $ c $ is the speed of light, and $\eta = m_\p m_\s / \m^2$.
To this order, the dominant relativistic secular contribution is on the
precession of the periapsis, with (Eq.\,\ref{100119b}):
\be
\dot \vv{e}_\b = \frac{3 \mu_\b n_\b}{c^2 a_\b (1-e_\b^2)} \, \vv{\hat \K}_\b
\times \vv{e}_\b \ .
\llabel{101029b}
\ee

\subsection{Spin motion}

\begin{figure}
\begin{center}
\includegraphics[width=\textwidth]{\figpath 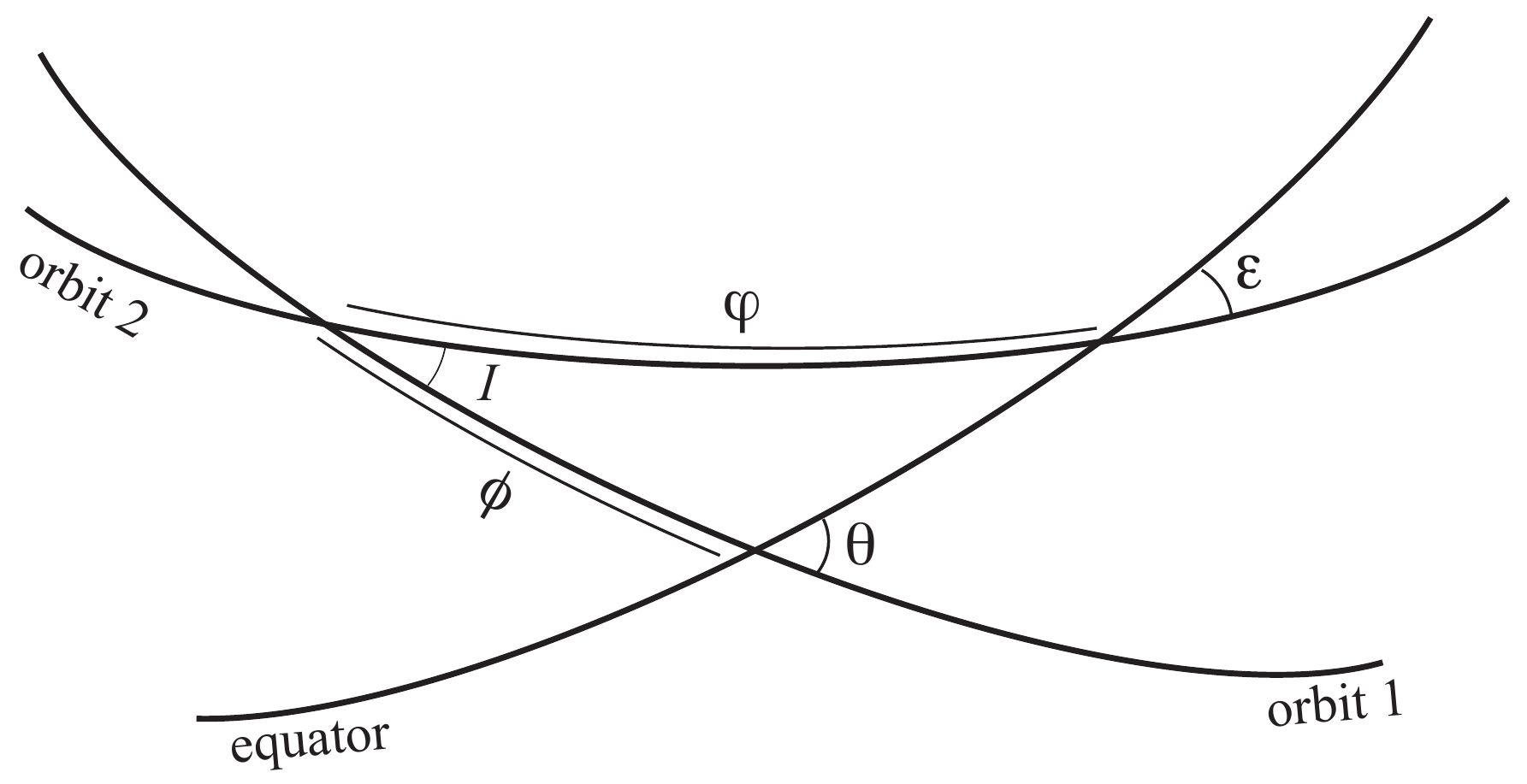} 
 \caption{Reference planes for the definition of the direction cosines and the precession angles.  \llabel{figspin}  }
\end{center}
\end{figure}

All bodies are oblate ellipsoids, so we need to take into account the deformation of the gravity field given by their structure.
The additional contribution to the potential energy (Eq.\,\ref{090514a}) is given in Jacobi coordinates by (see appendix~\ref{apenB}):

\begin{eqnarray}
U_R & =&  
G \frac{m_\p m_\s}{r_\b} \sum_{\ij=\p,\s} J_{2,\ij}
 \left(\frac{R_\ij}{r_\b}\right)^2 \!\! P_2 (\vv{\hat r}_\b \cdot \vv{\hat \k}_\ij) \nonumber \\ &+&
G \frac{\M \m}{r_\a} J_{2,\e} 
\left(\frac{R_\e}{r_\a}\right)^2 \!\! P_2 (\vv{\hat r}_\a \cdot \vv{\hat \k}_\e)
\ , \llabel{130116a}
\end{eqnarray}
where terms in $(R_\ij/r_j)^3$, $(R_\e/r_\a)^2 (r_\b / r_\a)^2$, and $m_\e / m_\ij (r_\b/r_\a)^3$ have been neglected ($\ij=0,1,2$, $j=1,2$).
As in previous section, we can obtain the contributions to the orbital motion
directly from equations (\ref{090514d}) and (\ref{100119b}) using $U_R$ instead of $U'$,

The evolution of the spins can be tracked by the rotational angular momenta, $ \vv{L}_\ij \simeq C_\ij \om_\ij \, \vv{\hat \k}_\ij $, where  $C_\ij$ are  the principal moments of inertia.
In Jacobi coordinates, the orbital angular momentum is equal to $\vv{G}_\b + \vv{G}_\a $ \citep{Poincare_1905}. 
Since the total angular momentum is conserved, the contributions to the spin of the bodies can be computed from their orbital contributions: 

\be 
\vv{\dot L}_\p + \vv{\dot L}_\s = - \vv{\dot G}_\b \ ,  \quad
\vv{\dot L}_\e = - \vv{\dot G}_\a \ .
\llabel{090514e}
\ee

Then, averaging again the equations of motion over the mean anomalies of both orbits (see appendix~\ref{apenA}) we get:

\begin{equation}
\vv{\dot L}_\p = 
-  \alpha_\p \cos \de_\p \, \vv{\hat \K}_\b \times \vv{\hat \k}_\p 
\ , \llabel{090514f}
\end{equation}

\begin{equation}
\vv{\dot L}_\s = 
-  \alpha_\s \cos \de_\s \, \vv{\hat \K}_\b \times \vv{\hat \k}_\s 
\ , \llabel{130117b}
\end{equation}

\begin{equation}
\vv{\dot L}_\e = 
- \alpha_\e \cos \ve_\e \, \vv{\hat \K}_\a \times \vv{\hat \k}_\e 
\ , \llabel{120116a}
\end{equation}

\begin{eqnarray}
\vv{\dot G}_\b = 
- \alpha_\p \cos \de_\p \, \vv{\hat \k}_\p \times \vv{\hat \K}_\b 
- \alpha_\s \cos \de_\s \, \vv{\hat \k}_\s \times \vv{\hat \K}_\b 
 \ , \llabel{090514z2s}
\end{eqnarray}
\begin{eqnarray}
\vv{\dot G}_\a = - \alpha_\e \cos \ve_\e \, \vv{\hat \k}_\e \times \vv{\hat \K}_\a 
 \ , \llabel{090514z1s}
\end{eqnarray}
\begin{eqnarray}
\vv{\dot e}_\b &=&
- \sum_{\ij=0,1} \Frac{\alpha_\ij}{\norm{\vv{G}_\b}} \Big[ \cos \de_\ij \,
\vv{\hat \k}_\ij \times \vv{e}_\b + \frac{1}{2} (1 - 5 \cos^2 \de_\ij) \,
\vv{\hat \K}_\b \times \vv{e}_\b \Big] \ , \llabel{120117fa}
\end{eqnarray}
\begin{eqnarray}
\vv{\dot e}_\a &=& - \Frac{\alpha_\e}{\norm{\vv{G}_\a}} \Big[ \cos \ve_\e \,
\vv{\hat \k}_\e \times \vv{e}_\a + \frac{1}{2} (1 - 5 \cos^2 \ve_\e) \,
\vv{\hat \K}_\a \times \vv{e}_\a \Big] \ , \llabel{120117fb}
\end{eqnarray}
where
\be 
\alpha_\p =  \frac{3 G m_\p m_\s J_{2,\p} R_\p^2}{2 a_\b^3 (1-e_\b^2)^{3/2}} \ ,
\llabel{090514h}
\ee
\be 
\alpha_\s =  \frac{3 G m_\p m_\s J_{2,\s} R_\s^2}{2 a_\b^3 (1-e_\b^2)^{3/2}} \ ,
\llabel{090514g}
\ee
\be 
\alpha_\e = \frac{3 G \M \m J_{2,\e} R_\e^2}{2 a_\a^3 (1-e_\a^2)^{3/2}} \ ,
\llabel{120116b}
\ee
and
\be
\cos \de_\ij = \vv{\hat \k}_\ij \cdot \vv{\hat \K}_\b \ , \quad
\cos \ve_\ij = \vv{\hat \k}_\ij \cdot \vv{\hat \K}_\a \ , \llabel{090514j}
\ee
are the direction cosines of the spins: $ \de_\ij$ is the obliquity
to the orbital plane of the inner orbit, and $\ve_\ij$ is the obliquity to the orbital
plane of the outer companion.
Using the mutual inclination between orbital planes (Eq.\,\ref{130117a}), the three direction cosines can also be related as
\be
\cos \ve_\ij = \cos I \cos \de_\ij + \sin I \sin \de_\ij \cos \phi_\ij 
\ , \llabel{090521a}
\ee
or
\be
\cos \de_\ij = \cos I \cos \ve_\ij - \sin I \sin \ve_\ij \cos \varphi_\ij \ ,
\llabel{160608a}
\ee
where $ \phi_\ij $ is the precession angle between the projections of $
\vv{\hat \k}_\ij $ and $ \vv{\hat \K}_\a $ in the plane normal to $ \vv{\hat \K}_\b $,
and $ \varphi_\ij $ is the precession angle between the projections of $
\vv{\hat \k}_\ij $ and $ \vv{\hat \K}_\b $ in the plane normal to $ \vv{\hat \K}_\a $ 
(see Fig.\,\ref{figspin}).

\subsection{Tidal effects}
\llabel{tidaleffects}

Neglecting the tidal effects raised by the planet in each star ($\ij = 0,1$), that is,  neglecting terms in $m_\e / m_\ij (r_\b/r_\a)^3$, the tidal potential energy of the system is given in Jacobi coordinates by (see appendix~\ref{apenC}):

\be
U_T = - \frac{G}{r_\b^3} \sum_{\ij=0,1} k_{2,\ij} \frac{m_\ji^2 R_\ij^5}{r_{\b \ij}'^3} P_2 (\vv{\hat r}_\b \cdot \vv{\hat r}_{\b \ij}') - k_{2,\e} \frac{G \m^2 R_\e^5}{r_\a^3 r_\a'^3} P_2 (\vv{\hat r}_\a \cdot \vv{\hat r}_\a')
\ , \llabel{130122e}
\ee
where $\ji = 1 - \ij$, 
and $\vv{r}_\ij'$ is the position of the interacting mass at a time delayed of $\Delta t_\ij$, which corresponds to the deformation time-lag of the body $\ij$. 
The dissipation of the mechanical energy of tides in the body's interior is responsible for this delay between the initial perturbation and the maximal deformation.
As the rheology of stars and planets is badly known, the exact dependence of $\Delta
t_\ij$ on the tidal frequency is unknown.
Many different authors have studied the problem and several models have been
developed so far, from the simplest ones to the more complex \citep[for a review
see][]{Efroimsky_Williams_2009, Ferraz-Mello_2013, Correia_etal_2014}.
The huge problem in validating one model better than others is the difficulty to
compare the theoretical results with the observations, as the effect of tides
are very small and can only be detected efficiently after long periods of time.
Nevertheless, for gaseous planets and stars, the amount of tidal energy that is dissipated in a cycle is rather small \citep[e.g.][]{Lainey_etal_2009, Lainey_etal_2012, Penev_etal_2012}, which is equivalent to short time responses.
In those cases, most viscoelastic tidal models can be made linear (weak friction), with a constant $\Delta t_\ij$ value \citep[see][]{Correia_etal_2014, Makarov_2015}.
Therefore, for simplicity, we adopt here a weak friction model with constant $\Delta t_\ij$
\citep[][]{Singer_1968, Alexander_1973, Mignard_1979}, for which:

\be
\vv{r}_{\b \ij}' \simeq \vv{r}_\b + \Delta t_\ij \left(\om_\ij \vv{\hat \k}_\ij
\times \vv{r}_\b - \vv{\dot r}_\b \right) \ , \quad
\vv{r}_\a' \simeq \vv{r}_\a + \Delta t_\e \left(\om_\e \vv{\hat \k}_\e
\times \vv{r}_\a - \vv{\dot r}_\a \right) \ . \llabel{090514c}
\ee

As for the spin motion, we can obtain the equations of motion directly from equations (\ref{090514d}), (\ref{100119b}) and (\ref{090514e}) using $U_T$ instead of $U'$, and then averaging over the mean anomalies of both orbits (see appendix~\ref{apenA}):

\be 
\vv{\dot G}_\b = - \sum_{\ij=0,1} \vv{\dot L}_\ij  \ ,  \quad
\vv{\dot G}_\a = - \vv{\dot L}_\e\ .
\llabel{110503a}
\ee

\begin{eqnarray}
\vv{\dot L}_\ij &=& K_\ij \,
n_\b \left[ f_4(e_\b) \sqrt{1-e_\b^2} \frac{\om_\ij}{2 n_\b} (\vv{\hat
\k}_\ij - \cos \de_\ij \, \vv{\hat \K}_\b) 
\right. \\ && \left.
- f_1(e_\b) \frac{\om_\ij}{n_\b} \vv{\hat \k}_\ij + f_2(e_\b)
\vv{\hat \K}_\b 
+  \frac{(\vv{e}_\b \cdot \vv{\hat \k}_\ij) (6 + e_\b^2)}{4
(1-e_\b^2)^{9/2}} \frac{\om_\ij}{n_\b}  \vv{e}_\b
\right] \ , \nonumber \llabel{090514knew}
\end{eqnarray}
\begin{eqnarray}
\vv{\dot L}_\e &=& K_\e \,
n_\a \left[ f_4(e_\a) \sqrt{1-e_\a^2} \frac{\om_\e}{2 n_\a} (\vv{\hat
\k}_\e - \cos \ve_\e \, \vv{\hat \K}_\a) 
\right. \\ && \left.
- f_1(e_\a) \frac{\om_\e}{n_\a} \vv{\hat \k}_\e + f_2(e_\a)
\vv{\hat \K}_\a 
+  \frac{(\vv{e}_\a \cdot \vv{\hat \k}_\e) (6 + e_\a^2)}{4
(1-e_\a^2)^{9/2}} \frac{\om_\e}{n_\a}  \vv{e}_\a
\right] \ , \nonumber \llabel{090514knew2}
\end{eqnarray}

\begin{eqnarray}
\dot \vv{e}_\b &=& 
\sum_{\ij=0,1} \frac{15}{2} \, k_{2,\ij} \, n_\b \left(
\Frac{m_\ji}{m_\ij} \right) \left( \Frac{R_\ij}{a_\b} \right)^5 f_4 (e_\b) 
\, \vv{\hat \K}_\b \times \vv{e}_\b \crm 
&-& \sum_{\ij=0,1} \Frac{K_\ij}{\beta_\b a_\b^2} 
\left[ f_4(e_\b) \frac{\om_\ij}{2 n_\b} (\vv{e}_\b \cdot \vv{\hat \k}_\ij) \,
\vv{\hat \K}_\b 
- \left( \frac{11}{2} f_4(e_\b) \cos \de_\ij \frac{\om_\ij}{n_\b} 
- 9 f_5(e_\b) \right) \vv{e}_\b \right]  \ , \llabel{100119h}
\end{eqnarray}
\begin{eqnarray}
\dot \vv{e}_\a &=& 
\frac{15}{2} \, k_{2,\e} \, n_\a \left(
\Frac{\m}{m_\e} \right) \left( \Frac{R_\e}{a_\a} \right)^5 f_4 (e_\a) 
\, \vv{\hat \K}_\a \times \vv{e}_\a \crm 
&-&  \Frac{K_\e}{\beta_\a a_\a^2} 
\left[ f_4(e_\a) \frac{\om_\e}{2 n_\a} (\vv{e}_\a \cdot \vv{\hat \k}_\e) \,
\vv{\hat \K}_\a 
- \left( \frac{11}{2} f_4(e_\a) \cos \ve_\e \frac{\om_\e}{n_\a} 
- 9 f_5(e_\a) \right) \vv{e}_\a \right]  \ , \llabel{100119h2}
\end{eqnarray}
where
\be
K_\ij = \Delta t_\ij \frac{3 k_{2,\ij} G m_\ji^2 R_\ij^5}{a_\b^6} \ , \quad
K_\e = \Delta t_\e \frac{3 k_{2,\e} G \m^2 R_\e^5}{a_\a^6} \ , \llabel{090514m}
\ee
and
\be
f_1(e) = \frac{1 + 3e^2 + 3e^4/8}{(1-e^2)^{9/2}} \ , \llabel{090514n}
\ee
\be
f_2(e) = \frac{1 + 15e^2/2 + 45e^4/8 + 5e^6/16}{(1-e^2)^{6}} \ , \llabel{090514o}
\ee
\be
f_3(e) = \frac{1 + 31e^2/2 + 255e^4/8 + 185e^6/16 + 25e^8/64}{(1-e^2)^{15/2}} \ , 
\llabel{090514p}
\ee
\be
f_4(e) = \frac{1 + 3e^2/2 + e^4/8}{(1-e^2)^5} \ , \llabel{090515d}
\ee
\be
f_5(e) = \frac{1 + 15e^2/4 + 15e^4/8 + 5e^6/64}{(1-e^2)^{13/2}} \ . \llabel{090515e}
\ee

The first term in expressions (\ref{100119h}) and (\ref{100119h2}) corresponds to the permanent tidal deformation, while the second term corresponds to the dissipative contribution.
%

\section{Secular dynamics}
\llabel{secevol}

In this section we look at the orbital and spin dynamics without tidal dissipation.
We first consider only the orbital dynamics without spin (section~\ref{orbdyn}) and then add to the equations of motion for the spin of a single body (section~\ref{spindyn}).
In some special configurations the secular equations of motion become integrable, which allow us to easily compute the possible trajectories for the system given its initial conditions.
Moreover, the equilibrium configurations for the orbit and spin, that correspond to the final outcomes of tidal evolution, become easy to identify.
For that purpose, we will use a reference hypothetical circumbinary system (Table\,\ref{standard}), which is similar to Kepler-34 \citep{Welsh_etal_2012}. 
In section~\ref{tidalevol}, this particular choice of parameters will allow us to illustrate some interesting dynamical effects that were not yet described in the literature.

\begin{table}
\caption{
Initial parameters for the Kepler-16 system \citep{Doyle_etal_2011, Winn_etal_2011} and for a hypothetical circumbinary system (that we call ``standard''), very similar to the Kepler-34 system \citep{Welsh_etal_2012}. The main differences between Kepler-34 and the standard system
are in the mass of star B, and in the semi-major axis of the inner orbit ($m_\s \approx M_\odot$ and $a_\b \approx 0.2$~AU for Kepler-34), in order to enhance tidal effects. The unknown parameters are compatible with observations for the Sun and giant planets in the Solar System. \llabel{standard} } 
\begin{center}
\begin{tabular}{|l|c|c|c|c|c|c|c|c|c|c|c|c|} \hline
   & \multicolumn{6}{|c|}{Kepler-16} & \multicolumn{6}{|c|}{standard} \\ \hline
parameter 
& \multicolumn{2}{|c|}{star A} & \multicolumn{2}{|c|}{star B} & \multicolumn{2}{|c|}{planet} 
& \multicolumn{2}{|c|}{star A} & \multicolumn{2}{|c|}{star B} & \multicolumn{2}{|c|}{planet} \\ \hline
$m$ $(M_\odot)$
& \multicolumn{2}{|c|}{$ 0.69 $} & \multicolumn{2}{|c|}{$ 0.20 $} & \multicolumn{2}{|c|}{$ 0.0003 $}  
& \multicolumn{2}{|c|}{$ 1.0 $} & \multicolumn{2}{|c|}{$ 0.2 $} & \multicolumn{2}{|c|}{$ 0.001 $} \\ 
$P_\mathrm{rot}$ $(\mathrm{day})$
& \multicolumn{2}{|c|}{$ 35.1 $} & \multicolumn{2}{|c|}{$ 20. $} & \multicolumn{2}{|c|}{$0.5$}  
& \multicolumn{2}{|c|}{$ 10. $} & \multicolumn{2}{|c|}{$ 1.0 $} & \multicolumn{2}{|c|}{$0.5$}  \\ 
$\de$ $(\mathrm{deg})$ 
& \multicolumn{2}{|c|}{$ 1.6 $} & \multicolumn{2}{|c|}{$ 10. $} & \multicolumn{2}{|c|}{$1.$}   
& \multicolumn{2}{|c|}{$ 5. $} & \multicolumn{2}{|c|}{$ 10. $} & \multicolumn{2}{|c|}{$20.$}  \\ 
$\phi$ $(\mathrm{deg})$ 
& \multicolumn{2}{|c|}{$ 0. $} & \multicolumn{2}{|c|}{$ 0. $} & \multicolumn{2}{|c|}{$0.$} 
& \multicolumn{2}{|c|}{$ 0. $} & \multicolumn{2}{|c|}{$ 0. $} & \multicolumn{2}{|c|}{$0.$}  \\ \hline
$R$ $(\times 10^6 \mathrm{m})$ 
& \multicolumn{2}{|c|}{$ 452. $} & \multicolumn{2}{|c|}{$ 157. $} & \multicolumn{2}{|c|}{$53.$}  
& \multicolumn{2}{|c|}{$ 695. $} & \multicolumn{2}{|c|}{$ 150. $} & \multicolumn{2}{|c|}{$70.$}  \\ 
$C / (m R^2)$ 
& \multicolumn{2}{|c|}{$ 0.08 $} & \multicolumn{2}{|c|}{$ 0.08 $} & \multicolumn{2}{|c|}{$0.20$} 
& \multicolumn{2}{|c|}{$ 0.08 $} & \multicolumn{2}{|c|}{$ 0.08 $} & \multicolumn{2}{|c|}{$0.25$}  \\ 
$k_2$ 
& \multicolumn{2}{|c|}{$ 0.028 $} & \multicolumn{2}{|c|}{$ 0.028 $} & \multicolumn{2}{|c|}{$0.50$} 
& \multicolumn{2}{|c|}{$ 0.028 $} & \multicolumn{2}{|c|}{$ 0.028 $} & \multicolumn{2}{|c|}{$0.50$}  \\ 
$\Delta t$ $(\mathrm{s})$ 
& \multicolumn{2}{|c|}{$ 0.1 $} & \multicolumn{2}{|c|}{$ 0.1 $} & \multicolumn{2}{|c|}{$100.$}  
& \multicolumn{2}{|c|}{$ 0.1 $} & \multicolumn{2}{|c|}{$ 0.1 $} & \multicolumn{2}{|c|}{$100.$}  \\ \hline
parameter 
& \multicolumn{3}{|c|}{\quad orbit $\b$ \quad \quad} & \multicolumn{3}{|c|}{orbit $\a$} 
& \multicolumn{3}{|c|}{\quad orbit $\b$ \quad \quad} & \multicolumn{3}{|c|}{orbit $\a$} \\ \hline
$a$ $(\mathrm{AU})$
& \multicolumn{3}{|c|}{$ 0.22 $} & \multicolumn{3}{|c|}{$ 0.70 $}  
& \multicolumn{3}{|c|}{$ 0.1 $} & \multicolumn{3}{|c|}{$ 1.5 $}  \\ 
$e$
& \multicolumn{3}{|c|}{$ 0.16 $} & \multicolumn{3}{|c|}{$ 0.01 $} 
& \multicolumn{3}{|c|}{$ 0.5 $} & \multicolumn{3}{|c|}{$ 0.2 $}  \\ 
$\varpi$ $(\mathrm{deg})$ 
& \multicolumn{3}{|c|}{$263.$} & \multicolumn{3}{|c|}{$318.$} 
& \multicolumn{3}{|c|}{$0.$} & \multicolumn{3}{|c|}{$180.$}  \\ \hline
$I$ $(\mathrm{deg})$ 
&\multicolumn{6}{|c|}{$1.$} 
&\multicolumn{6}{|c|}{$10.$}  \\  \hline
\end{tabular}
\end{center}
\end{table}

\subsection{Orbital dynamics}
\llabel{orbdyn}

The non-resonant dynamics of circumbinary planets has been studied in great detail by \citet{Migaszewski_Gozdziewski_2011}.
Here we recall some of the main features of the problem, in particular those corresponding to an integrable problem, which will allow us to understand the more complex dynamics when the effect from the spins and tides are included.

\subsubsection{Coplanar systems}

When $\sin I=0$ (coplanar systems), the average of the orbital energy over the mean anomalies, in the octupolar approximation (Eq.\,\ref{090514a}) is given by \citep[e.g.][]{Lee_Peale_2003, Laskar_Boue_2010, Correia_etal_2012}
\be
\moy{U} =  
- \frac{\gamma_2}{3} (1 + \frac{3}{2} e_\b^2)
+ 4 \gamma_3 (1 + \frac{3}{4} e_\b^2) e_\b e_\a \cos \Delta \varpi
\ , \llabel{090514am}
\ee
where $\Delta \varpi = \varpi_\b - \varpi_\a$, and $\varpi_\ij$ is the \bfx{longitude of the pericenter} of each orbit.
\bfx{The parameters $\gamma_2$ and $\gamma_3$ are given by expressions (\ref{090514i}) and (\ref{120117a}), respectively, and depend only on $e_\a$.}
For very eccentric inner orbits, the above orbital energy can be corrected for general relativity effects (Eq.\,\ref{160503a}) through the additional contribution \citep[e.g.][]{Touma_etal_2009}:
\be
\moy{U_\mathrm{gr}} = - \frac{3 \beta_\b \mu_1^2}{a_\b^2 c^2 (1-e_\b^2)^{1/2}} \ .
\llabel{relats}
\ee
The secular system (\ref{090514am},\ref{relats}) does not depend on the mean longitudes, so the semi-major axes are constant. Moreover, it depends on a unique angle $\Delta \varpi$. It is thus integrable. 
The eccentricities are related through the total orbital angular momentum (Eq.\,\ref{130116c}):
\be
 \beta_\b \sqrt{\mu_\b a_\b (1-e_\b^2)} +  \sigma \beta_\a \sqrt{\mu_\a a_\a (1-e_\a^2)}  = C=cte 
\ . \llabel{150412a}
\ee
where $\sigma=\pm 1$, depending whether the orbits rotate in the same way or with opposite direction. Here we will only consider the 
$\sigma=+1$ case of planets orbiting in the same direction. 

\begin{figure}
\begin{center}
\includegraphics[width=\textwidth]{\figpath 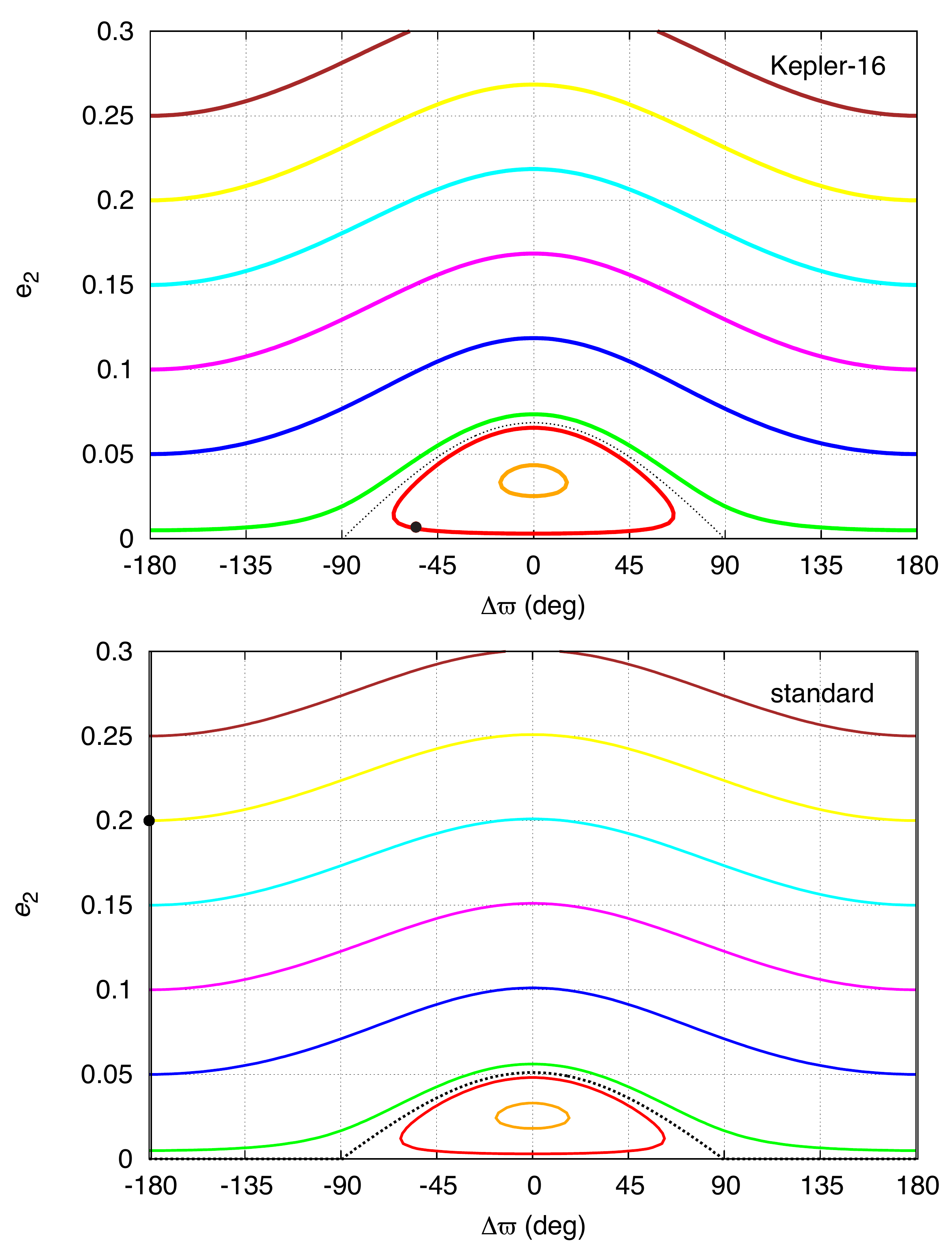} 
 \caption{Level curves for the outer orbit eccentricity of Kepler-16 and the standard system (Table\,\ref{standard}). There are essentially two possible kinematic regimes: oscillation around $\Delta \varpi = 0^\circ$ and circulation. The present position of each system is given by a dot.
 \llabel{figocto}  }
\end{center}
\end{figure}

In Figure~\ref{figocto} we plot the level curves of the total energy in the plane $(e_\a, \Delta \varpi)$ for Kepler-16 and for the standard system with $I=0$ (Table~\ref{standard}).
There are essentially two possible \bfx{kinematic regimes for $\Delta \varpi$ in Figure~\ref{figocto}: oscillation around $ 0^\circ$ or circulation between $-180^\circ$ and $180^\circ$.}
The standard system (Table\,\ref{standard}) is in circulation, but for Kepler-16 we cannot be sure.
The best fit data \citep{Doyle_etal_2011} gives $\Delta \varpi \approx (-55\pm20)^\circ$.
For the mean value of this estimate, the system is oscillating around $0^\circ$, but for the lower bound  $\Delta \varpi \approx -75^\circ$ it is very near circulation zone (the dotted line gives the transition of kinematic regime).
\bfx{Although the angle $\Delta \varpi$ may present a different behavior, the dynamics in the plane ($e_\a \sin \Delta \varpi, e_\a \cos \Delta \varpi$) always correspond to circulation of all trajectories around a fix point ($e_2 \approx 0.033, \Delta \varpi = 0^\circ$).
For both systems}, we observe that the eccentricity of the outer orbit only undergoes small variations.
In the secular system, as the semi-major axes are constant, the difference of the angular momentum (\ref{150412a}) with respect to the circular angular momentum, i.e. the AMD \citep{Laskar_2000} is also constant, that is 
\be
 \beta_\b \sqrt{\mu_\b a_\b} (1-\sqrt{1-e_\b^2}) +   \beta_\a \sqrt{\mu_\a a_\a} (1- \sqrt{1-e_\a^2})  = D=cte 
\ . \llabel{AMDp}
\ee
Thus, with $\chi(e)=1-\sqrt{1-e^2}$, 
\be
  \chi(e_\b) = \Frac{D}{\beta_\b \sqrt{\mu_\b a_\b}} - \Frac{\beta_\a \sqrt{\mu_\a a_\a}}{\beta_\b \sqrt{\mu_\b a_\b}}\chi(e_\a)
\ee
Since most of the orbital angular momentum is contained in the inner orbit (${\beta_\a \sqrt{\mu_\a a_\a}}\ll{\beta_\b \sqrt{\mu_\b a_\b}}$), the variations in $\chi(e_\b)$ (resp. $e_\b$)  are  smaller than those in $\chi(e_\a)$ (resp. $e_\a$), so the inner orbit eccentricity $e_\b$ can be considered almost as constant.
For Kepler-16 and the standard systems we have $m_\p \gg m_\s$, but for other circumbinary systems, such as Kepler-34 or Kepler-35 \citep{Welsh_etal_2012}, we have $m_\p \sim m_\s$.
In those cases $\gamma_3 \approx 0$, so the octupolar contribution vanishes (Eq.\,\ref{090514am}), and both eccentricities remain almost constant.

\subsubsection{Inclined systems}

\begin{figure}
\begin{center}
\includegraphics[width=\textwidth]{\figpath 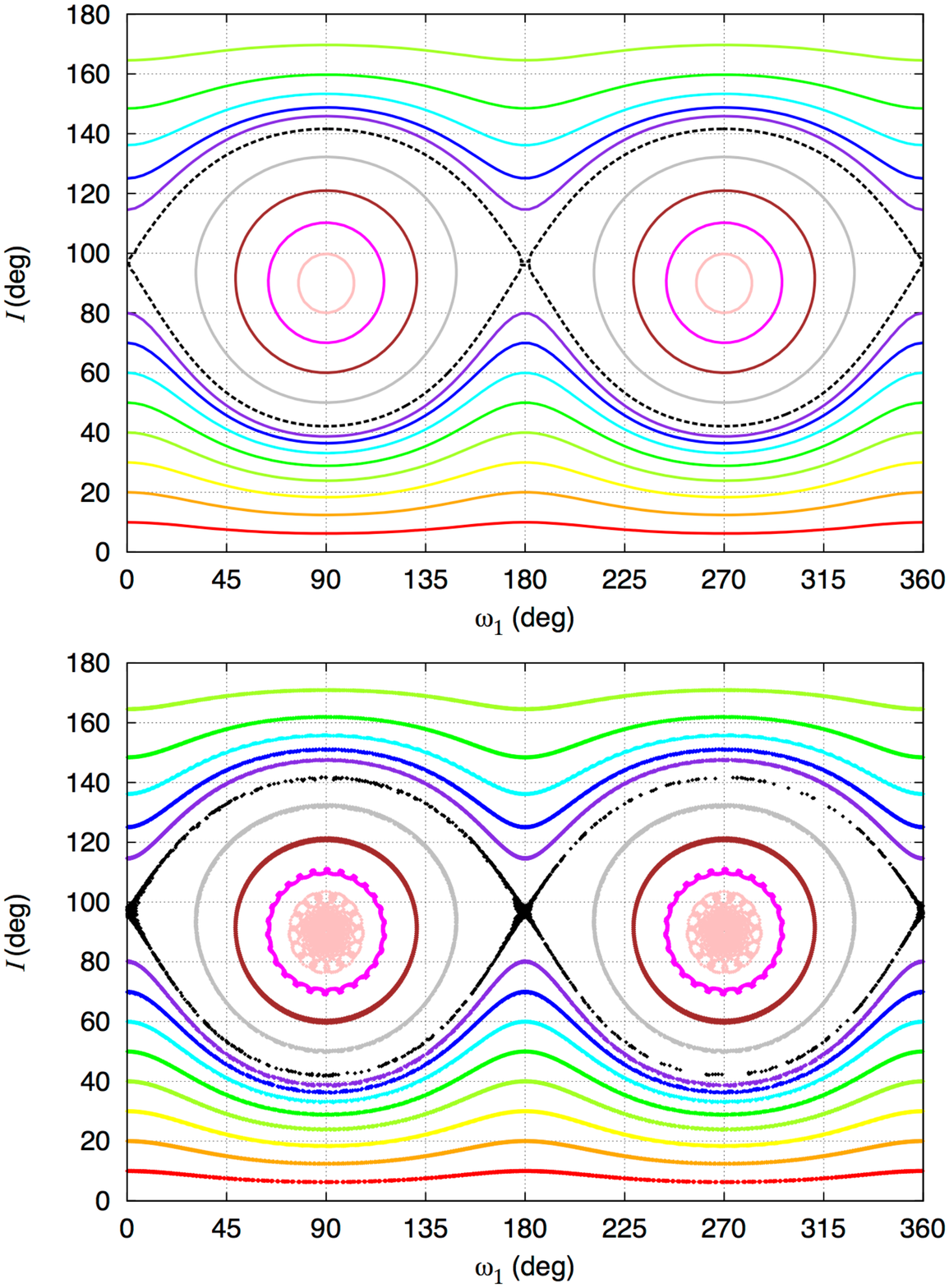} 
 \caption{Secular trajectories for the standard system (Table\,\ref{standard}) seen in the plane $(I, \w_\b)$. We show the trajectories using the quadrupolar approximation (top), corresponding to the level curves of constant energy, and using the octupolar approximation (bottom), obtained with numerical simulations. There are two possible dynamical regimes: libration around $\w_\b = \pm 90^\circ$ and circulation \citep[see also][]{Migaszewski_Gozdziewski_2011}. 
 \llabel{figquadi}  }
\end{center}
\end{figure}

\begin{figure}
\begin{center}
\includegraphics[width=\textwidth]{\figpath 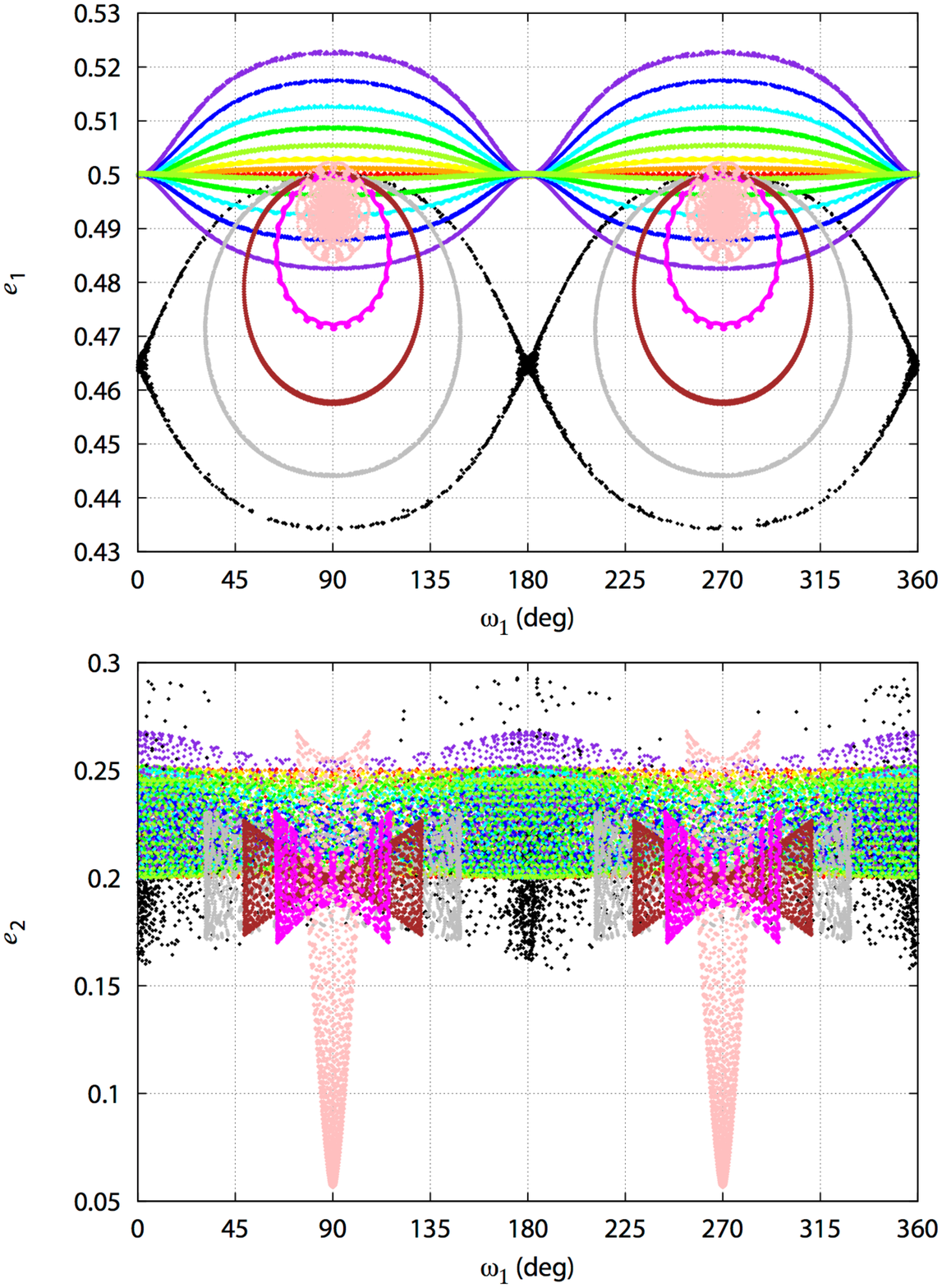} 
 \caption{Secular trajectories for the standard system (Table\,\ref{standard}) seen in the planes $(e_\b, \w_\b)$ and $(e_\a, \w_\b)$. We show the trajectories using the octupolar approximation, obtained with numerical simulations \citep[for more details see][]{Migaszewski_Gozdziewski_2011}. 
 \llabel{figquade}  }
\end{center}
\end{figure}

For inclined systems ($\sin I \ne 0 $), the average of the orbital energy (Eq.\,\ref{090514a}) over the mean anomalies additionally depends on $I$, $\varpi_\b$ and $\varpi_\a$ \citep[e.g.][]{Harrington_1968, Lidov_Ziglin_1976, Laskar_Boue_2010}, and the problem is no longer integrable. 
However, while for the coplanar problem the main dynamical features result from the octupolar contribution (term in $\Delta \w$, Eq.\,\ref{090514am}), for the inclined problem the major contribution comes from the quadrupolar term.
\bfx{Expressing this term in the Laplace invariable plane, for which $\varpi_\b = \w_\b$ and $\varpi_\a = \pi + \w_\a$, where $\w_\ij$ is the longitude of the pericenter of each orbit measured from the line of the nodes between the two orbits}, we get \citep{Correia_etal_2013}:
\be
\moy{U_\mathrm{qd}} = - \frac{\gamma_2}{3} \left[ (1 + \frac{3}{2} e_\b^2 ) (1-\frac{3}{2} \sin^2 I )+\frac{15}{4} e_\b^2 \sin^2 I  \cos 2 \w_\b \right]
\ . \llabel{090514aq}
\ee
Moreover, $\cos I$ (and thus $\sin I$)  can be expressed in terms of $G_\b, G_\a$ 
 through the conservation of the total orbital angular momentum (Eq.\,\ref{130116c})
\be
G_\b^2 + G_\a^2 + 2 G_\b G_\a \cos I =  G_{tot}^2 
\llabel{150413a}
\ee
%

Thus, if we restrict the dynamics to the quadrupolar approximation, 
here again, the Hamiltonian depends on a single angle $\w_\b$ and the problem is integrable
\citep{Harrington_1968, Lidov_Ziglin_1976}. 
Moreover, the outer orbit eccentricity $e_\a$ and $\gamma_2$ are constant (Eq.\ref{090514i}), since there is no contribution from $\w_\a$ \citep{Harrington_1968, Lidov_Ziglin_1976, Farago_Laskar_2010}.
Therefore, the orbital energy only depends on $e_\b$ and $\w_\b$.
For $\norm{G_\b} \ll \norm{G_\a}$ previous equation can be simplified to get $ \sqrt{(1-e_\b^2)} \cos I \approx cte$, which is at the origin of the Lidov-Kozai mechanism \citep{Lidov_1962, Kozai_1962, Lidov_Ziglin_1976}.
However, for circumbinary planets we expect $\norm{G_\b} > \norm{G_\a}$, so we cannot neglect the first term in expression (\ref{150413a}).

In Figure~\ref{figquadi} (top) we plot the level curves of the total energy in the plane $(I, \w_\b)$ for the standard system (Table~\ref{standard}), with different values for the inclination.
We observe that there are two possible dynamical regimes: circulation and libration around $\w_\b = \pm 90^\circ$ for high inclinations values.
In the example shown we have $40^\circ \lesssim I \lesssim 140^\circ$, but these boundaries strongly depend on the eccentricity of the inner orbit.
For small $e_\b$ values the libration zone is restricted to the vicinity of $I \sim 90^\circ$, while for large $e_\b$ values the libration region can reach very low inclinations \citep[see][]{Farago_Laskar_2010}.
Moreover, these boundaries are not completely symmetric, since the separatrix between the libration and circulation regimes is shifted towards $I > 90^\circ$.
This shift cannot be observed in the classic restricted case \citep[e.g.][]{Verrier_Evans_2009}, but in the planetary case with general relativity there is an increase in the precession rate of $\w_\b$ that breaks the symmetry.

In Figure~\ref{figquadi} (bottom) we plot the exact same trajectories performing numerical simulations using the full model form section~\ref{orbmotion} (octupole approximation).
We observe that the phase space is almost unchanged, there is only some additional chaotic diffusion for the trajectories near the separatrix.  
The same is valid when we plot these trajectories in the plane $(e_\b, \w_\b)$ as shown in Figure~\ref{figquade} (top).
Although the inclination may undergo significant variations, the eccentricity of the inner orbit only varies by a very small amount due to the conservation of the total angular momentum (Eq.\,\ref{150413a}).

We conclude that the quadrupolar approximation captures the main dynamical features for the $e_\b, I$ and $\w_\b$ parameters.
However, these conclusions cannot be extended to $e_\a$.
In the quadrupolar approximation the eccentricity of the outer orbit is constant, but in the octupolar approximation it undergoes some variations due to the presence of term in $\w_\a$ (Eq.\,\ref{090514am}).
For the coplanar case 
we have $0.2 \le e_\a \le 0.25$ (Fig.\,\ref{figocto}).
In Figure~\ref{figquade} (bottom) we show the same trajectories from Figure~\ref{figquadi} in the plane 
$(e_\a, \w_\b)$.
For trajectories in circulation far from the separatrix, i.e., for $\sin I \approx 0$, the outer orbit eccentricity is still bounded by $0.2 \le e_\a \le 0.25$, like in the coplanar case.
As the orbits become closer to the separatrix, the diffusion increases until a maximum $0.15 \le e_\a \le 0.3$.
For trajectories in libration far from the separatrix the eccentricity oscillations are bounded by $0.17 \le e_\a \le 0.23$, except for those close to the equilibrium points, for which the eccentricity undergoes large amplitude variations $0.05 \le e_\a \le 0.27$.

\subsection{Spin dynamics}
\llabel{spindyn}

\subsubsection{Planetary spin}

We now consider the spin of the planet into the analysis. 
Averaging the rotational energy (Eq.\,\ref{130116a}) over the mean anomalies gives \citep[e.g.][]{Goldreich_1966a}
\be
\moy{U_{R,\e}} = - \frac{\alpha_\e}{2} (\cos \ve_\e)^2 \ .
\llabel{150420a}
\ee
When we add the rotational energy to the orbital energy we get some additional degrees of freedom, so the problem is not integrable. 
However, in previous section we saw that the eccentricity of the inner orbit only undergoes small variations (Fig.\,\ref{figquade}).
Therefore, assuming $e_\b$ as constant, we can additionally average the orbital energy (Eq.\,\ref{090514aq}) over the argument of the pericenter, $\w_\b$.
Restricting the orbital contribution to the quadrupolar term, and suppressing the constant terms in $e_\b$,  gives for the total energy:
\be
U_\e = \moy{U_{R,\e}} + \moy{\moy{U_\mathrm{qd}}}_{\w_\b}
 = - \frac{\alpha_\e}{2} (\cos \ve_\e)^2 - \frac{\gamma_2}{2} (1 + \frac{3}{2} e_\b^2 ) \cos^2 I \ .
\llabel{150420b}
\ee
This reduced expression for the energy only depends on the direction cosines $\cos \ve_\e$ and $\cos I$, which give the relative directions of the angular momentum components, together with $\cos \de_\e$ (Eq.\,\ref{090514j}).
The three directions are related through the total angular momentum (with $L_i = \norm{\vv{L}_i}$) 
\be
L_\e G_\b \cos \de_\e + L_\e G_\a \cos \ve_\e + G_\b G_\a \cos I = K_\e = cte \ .
\llabel{150421a}
\ee
This problem is then integrable \citep{Boue_Laskar_2006, Boue_Fabrycky_2014b, Correia_2015}, since the three degrees of freedom (given by the direction cosines) can be related through the total energy (Eq.\,\ref{150420b}) and the total angular momentum (Eq.\,\ref{150421a}).
Moreover, we usually have $L_\e \ll G_\a \ll G_\b$, so previous expression can be simplified as
\begin{eqnarray}
\cos I &=& \cos I_0 - \frac{L_\e}{G_\a} \cos \de_\e  + O(\frac{L_\e}{G_\b})\ , \\
\cos^2 I &=& \cos^2 I_0 - 2\cos I_0\frac{L_\e}{G_\a} \cos \de_\e  + O(\frac{L_\e}{G_\b})\ , 
\llabel{150505b}
\end{eqnarray}
with
\be
\cos I_0 = \frac{K_\e}{G_\b G_\a} = cte \ .
\llabel{150505c}
\ee
We conclude that the mutual inclination is almost constant ($I \approx I_0$) and, leaving out the constant terms, 
  the total energy (Eq.\,\ref{150420b}) becomes:
\be
U_\e \approx - \frac{\alpha_\e}{2} (\cos \ve_\e)^2 - g_\e  \cos \de_\e \ ,
\llabel{150505d}
\ee
with
\be
g_\e =  -\gamma_2 (1 + \frac{3}{2} e_\b^2) \cos I_0 \frac{L_\e}{G_\a} = cte \ .
\llabel{150505e}
\ee
$g_\e / L_\e$ is the precession rate of the line of nodes of the two orbital planes.
The total energy expressed in this form is equivalent to the one obtained for previous works that study the spin evolution of a planet around a single star, whose orbit is perturbed by other planetary companions \citep[e.g.][]{Ward_Hamilton_2004}.

Planetary spins are often described with respect to their orbital plane.
In order to better understand the secular trajectories for the spin of the planet, we perform 
a  change of variables from polar to rectangular coordinates 
$(\ve_\e,\varphi_\e ) \longrightarrow (\uo, \vo)$ that gives the projection of the spin in the orbit \citep{Correia_2015}:
\be
\uo = \sin \ve_\e \cos \varphi_\e  
\quad ; \quad
\vo = \sin \ve_\e \sin \varphi_\e  \ ,
 \llabel{141230a}
\ee
where $\varphi_\e$ is the precession angle measured along the outer orbit from the inner orbit to the  equatorial plane of the planet (Fig.\,\ref{figspin}).
Thus,
\be
\cos \ve_\e = \sqrt{1 - \uo^2 - \vo^2} \ , 
\llabel{150505a}
\ee
and, with $I \approx I_0$, 
\be
\cos \de_\e = \cos I \cos \ve_\e - \sin I \sin \ve_\e \cos \varphi_\e \approx \cos I_0 \sqrt{1 - \uo^2 - \vo^2} - \uo \sin I_0 \ . 
\llabel{150421z}
\ee
Replacing the above expressions for the direction cosines in expression (\ref{150505d}) finally gives for the total energy:
\be
U_\e \approx \frac{\alpha_\e}{2} \left(\uo^2 + \vo^2 \right) - g_\e \left(\cos I_0 \sqrt{1 - \uo^2 - \vo^2} - \uo \sin I_0 \right) \ .
\llabel{150505f}
\ee

\begin{figure}
\begin{center}
\includegraphics[width=\textwidth]{\figpath 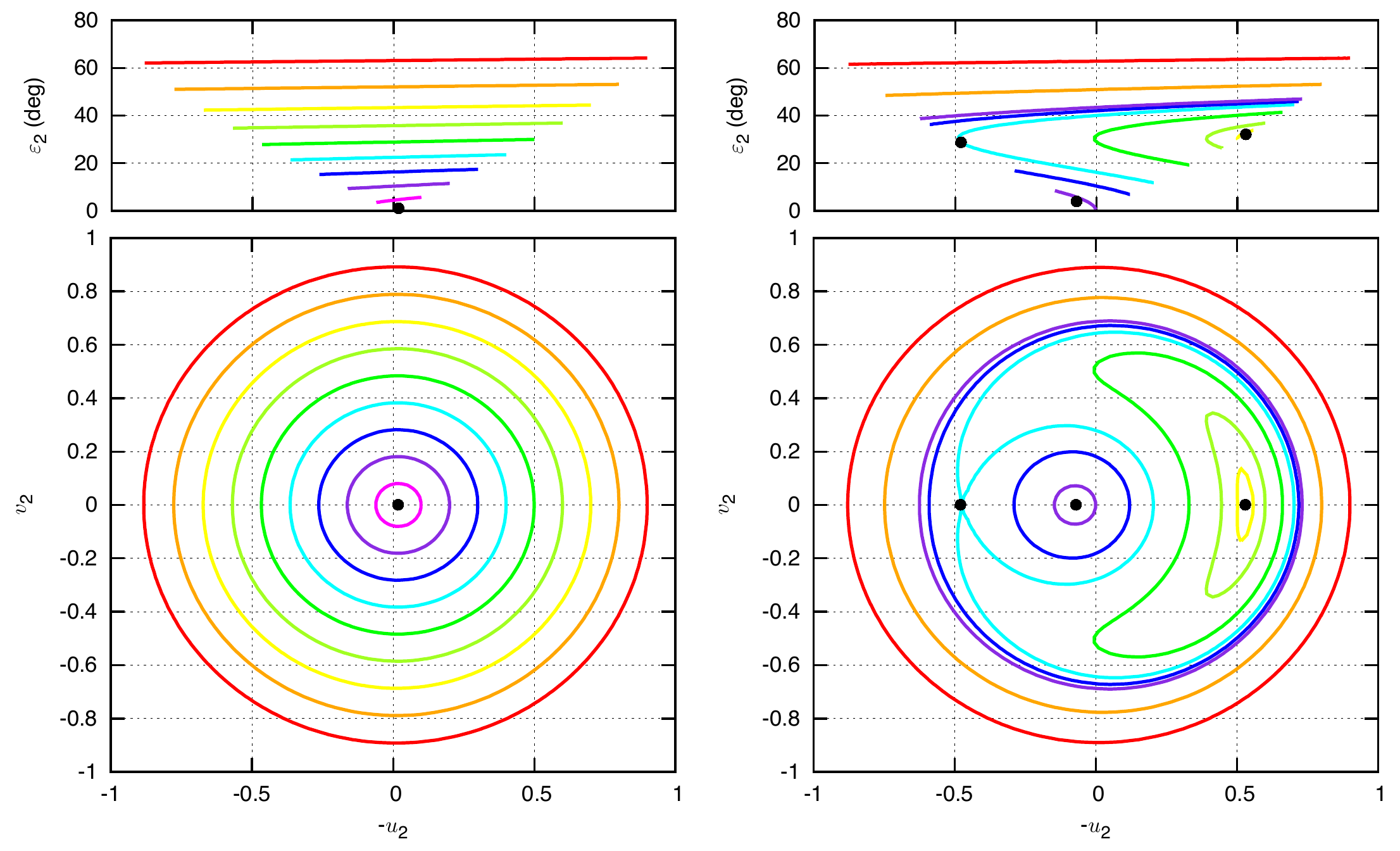} 
 \caption{Secular trajectories for the spin of Kepler-16\,b in the present system (left) and for a modified system with $m_\s = 0.8\,M_{Jup}$ (right). We show the spin projected on the orbit normal (top), and its projection on the orbital plane (bottom). Cassini states are marked with a dot.
 \llabel{figPspin}  }
\end{center}
\end{figure}

In Figure~\ref{figPspin}\,(left) we plot the level curves of the total energy in the plane $(\uo, \vo)$ for Kepler-16\,b \citep{Doyle_etal_2011}, adopting a rotation period of 0.5~day and $k_{2,\e} = 0.5$ to compute the $J_{2,\e}$ value for the planet (Eq.\,\ref{101220a}), and $C/(\M R_\e^2) = 0.2$ to compute the rotational angular momentum (Table~\ref{standard}).
Since $\alpha_\e \ll -g_\e$ and $I_0 \approx 1^\circ$, the total energy is dominated by the middle term $U_\e \approx   -g_\e \cos I_0 \sqrt{1 - \uo^2 - \vo^2}$ (Eq.\,\ref{150505f}).
We thus observe that the spin describes  almost circular trajectories with constant obliquity.
The equilibria for the spin can be obtained by finding the \bfx{critical points} of the total energy \citep{Correia_2015}:
\be
\frac{\partial U_\e}{\partial \uo} = 0 \quad ; \quad \frac{\partial U_\e}{\partial \vo} = 0 \ .
\llabel{150513a}
\ee 
These equations impose that $\vo = 0$ (which is equivalent to $\varphi_\e = 0$), meaning that the three angular momentum vectors lie in the same plane.
These equilibria are known in the literature as Cassini states \citep[e.g.][]{Colombo_1966, Ward_1975, Correia_2015}.
The first equation combined with $\vo = 0$ gives:
\be
\alpha_\e \, \uo + g_\e \left(\frac{\uo \cos I_0}{\sqrt{1 - \uo^2}} + \sin I_0 \right) = 0 \ ,
\llabel{150513b}
\ee 
where $\uo = \sin \ve_\e$ (Eq.\,\ref{141230a}).
This expression is equivalent to the commonly used condition to find the Cassini states equilibria \citep[e.g][]{Ward_Hamilton_2004}:
\be
\alpha_\e \sin \ve_\e \cos \ve_\e + g_\e \sin \left( \ve_\e + I_0 \right) = 0 \ .
\llabel{150513c}
\ee 
When $\alpha_\e \ll -g_\e$, there is one equilibrium point for $\ve_\e \approx - I_0$ (Fig.\,\ref{figPspin}\,left).
This is often the case for circumbinary planets, so we do not expect any significant oscillations in their obliquities.
In addition, as long as the system is nearly coplanar, the equilibrium obliquity remains very small.
However, for $\alpha_\e \approx g_2$, the phase space becomes much more interesting.
This situation occurs when the precession of the planet's spin and the precession of its orbit are near resonance
 ($\alpha_\e \cos \ve_\e / L_\e \approx \gamma_2 \cos I_0 / G_\a$), which can be obtained if Kepler-16's secondary is replaced by another Jupiter-mass planet. 
In Figure~\ref{figPspin}\,(right) we plot the level curves of the total energy for a modified Kepler-16 system, with $m_\s = 0.8\,M_{Jup}$.
In this case, high obliquity Cassini states are possible for the outer planet, similarly to what is observed for Saturn in the Solar System \citep{Ward_Hamilton_2004}.

\subsubsection{Stellar spins}

We now consider the effect of the spin of one of the stars in our analysis, for instance, the primary with mass $m_\p$.
As in previous section, we restrict the orbital contribution to the quadrupolar term, so that the problem remains integrable.
The total energy in this case is then given by: 
\be
U_\p = \moy{U_{R,\p}} + \moy{\moy{U_\mathrm{qd}}}_{\w_\b}
 = - \frac{\alpha_\p}{2} \cos^2 \de_\p - \frac{\gamma_2}{2} (1 + \frac{3}{2} e_\b^2 ) \cos^2 I \ .
\llabel{150513d}
\ee
This reduced expression for the energy now depends on the projection of the spin in the inner orbit
$\cos \de_\p$, but the problem remains integrable as the three direction cosines can be related through the total angular momentum,
\be
L_\p G_\b \cos \de_\p + L_\p G_\a \cos \ve_\p + G_\b G_\a \cos I = K_\p = cte \ .
\llabel{150513e} 
\ee
This expression is similar to expression (\ref{150421a}) from previous section, where the rotational angular momentum of the planet is replaced by that of the star.
However, the similarities with the planetary case end here, because for the star the three angular 
momenta may present similar magnitudes, i.e., $L_\p \sim G_\b \sim G_\a$.

In order to better understand the secular trajectories for the spin of the star, we can nevertheless 
perform a similar variable change $(\de_\p,\phi_\p ) \longrightarrow (\ui, \vi)$ that gives the projection of the spin of the star in the inner orbit:
\be
\ui = \sin \de_\p \cos \phi_\p  
\quad ; \quad
\vi = \sin \de_\p \sin \phi_\p  \ ,
 \llabel{150513f}
\ee
where $\phi_\p$ is the precession angle measured along the inner orbit from the outer orbit to the  equatorial plane of the star (Fig.\,\ref{figspin}).
Thus,
\be
\cos \de_\p = \sqrt{1 - \ui^2 - \vi^2} \ , 
\llabel{150513g}
\ee
and
\be
\cos \ve_\p = \cos I \cos \de_\p + \sin I \sin \de_\p \cos \phi_\p = \cos I \sqrt{1 - \ui^2 - \vi^2} + \ui \sin I \ . 
\llabel{150513h}
\ee
Replacing previous expression for $\cos \ve_\p$ in expression (\ref{150513e}) for the total angular momentum provides us an expression for $\cos I$ that depends only on the new variables ($\ui, \vi$),
which can be explicitly solved as \citep{Correia_2015}
\be
\cos I = \frac{(G_\b + L_\p \cos \de_\p) \Z - L_\p \ui \sqrt{1-\Z^2}}{\G}
\ , \llabel{141217f}
\ee
with
\be
\Z = \Z(\ui, \cos \de_\p) = \frac{K_\p - L_\p G_\b \cos \de_\p}{G_\a \G} 
\ , \llabel{141224a1}
\ee
and
\be
\G = \G(\ui, \cos \de_\p) = \sqrt{(G_\b + L_\p \cos \de_\p)^2+(L_\p \ui)^2}
\ . \llabel{141224a2}
\ee
Therefore, $\cos \de_\p$ and $\cos I$ depend only on the new variables ($\ui,\vi$), as well as the total energy given by expression (\ref{150513d}).
As for the planetary spin, the equilibria for the stellar spin are obtained from the \bfx{critical points} of the total energy: 
\be
\frac{\partial U_\p}{\partial \ui} = 0 \quad ; \quad \frac{\partial U_\p}{\partial \vi} = 0 \ .
\llabel{150513a2}
\ee 
Since 
$U_\p (\ui, \vi) = U_\p (\ui, \cos \de_\p(\ui,\vi))$, the second equation becomes
\be
\left.\frac{\partial U_\p}{\partial \vi}\right\vert_{\ui} = \left.\frac{\partial U_\p}{\partial (\cos \de_\p)}\right\vert_{\ui}
\left.\frac{\partial (\cos \de_\p)}{\partial \vi}\right\vert_{\ui} =
 - \left.\frac{\partial U_\p}{\partial (\cos \de_\p)}\right\vert_{\ui}\frac{\vi}{\cos \de_\p} = 0 \ .
\llabel{150514a}
\ee 
We hence conclude that $\vi = 0$ is still a possible solution (which is equivalent to $\phi_\p = 0$), meaning that the three angular momentum vectors lie again in the same plane.
The first equation combined with $\vi = 0$ thus gives a generalised version of Cassini states for the spin of the star \citep{Correia_2015}:
\be
\alpha_\p \, \ui + g (\ui) f (\ui) = 0
\ , \llabel{150514b}
\ee 
with $\cos \de_\p = \sqrt{1-\ui^2}$ ,
\be
g (\ui) = - \gamma_2 (1 + \frac{3}{2} e_\b^2 ) \cos I \frac{L_\p}{\G}
\ ,  \llabel{150514c}
\ee
\be
f(\ui) = \left(\frac{\ui \Z}{\sqrt{1-\Z^2}} + \frac{G_\b}{L_\p}+ \cos \de_\p \right) \left.\frac{\partial \Z}{\partial \ui}\right\vert_{\vi} 
 - \frac{\ui \Z}{\cos \de_\p} - \sqrt{1-\Z^2} -\frac{\cos I}{L_\p} \left.\frac{\partial \G}{\partial \ui}\right\vert_{\vi}  
\ , \llabel{150514d}
\ee
\be
\left.\frac{\partial \Z}{\partial \ui}\right\vert_{\vi}  = \frac{L_\p}{\G} \frac{G_\b}{G_\a} \frac{\ui}{\cos \de_\p} 
- \frac{\Z}{\G} \left.\frac{\partial \G}{\partial \ui} \right\vert_{\vi} 
\ , \quad \mathrm{and} \quad
\left.\frac{\partial \G}{\partial \ui}\right\vert_{\vi}  = - \frac{G_\b L_\p \ui}{\G \cos \de_\p} 
\ . \llabel{150514e}
\ee

\begin{sidewaysfigure}
\vskip12.5cm
\includegraphics[width=\textwidth,height=11cm]{\figpath 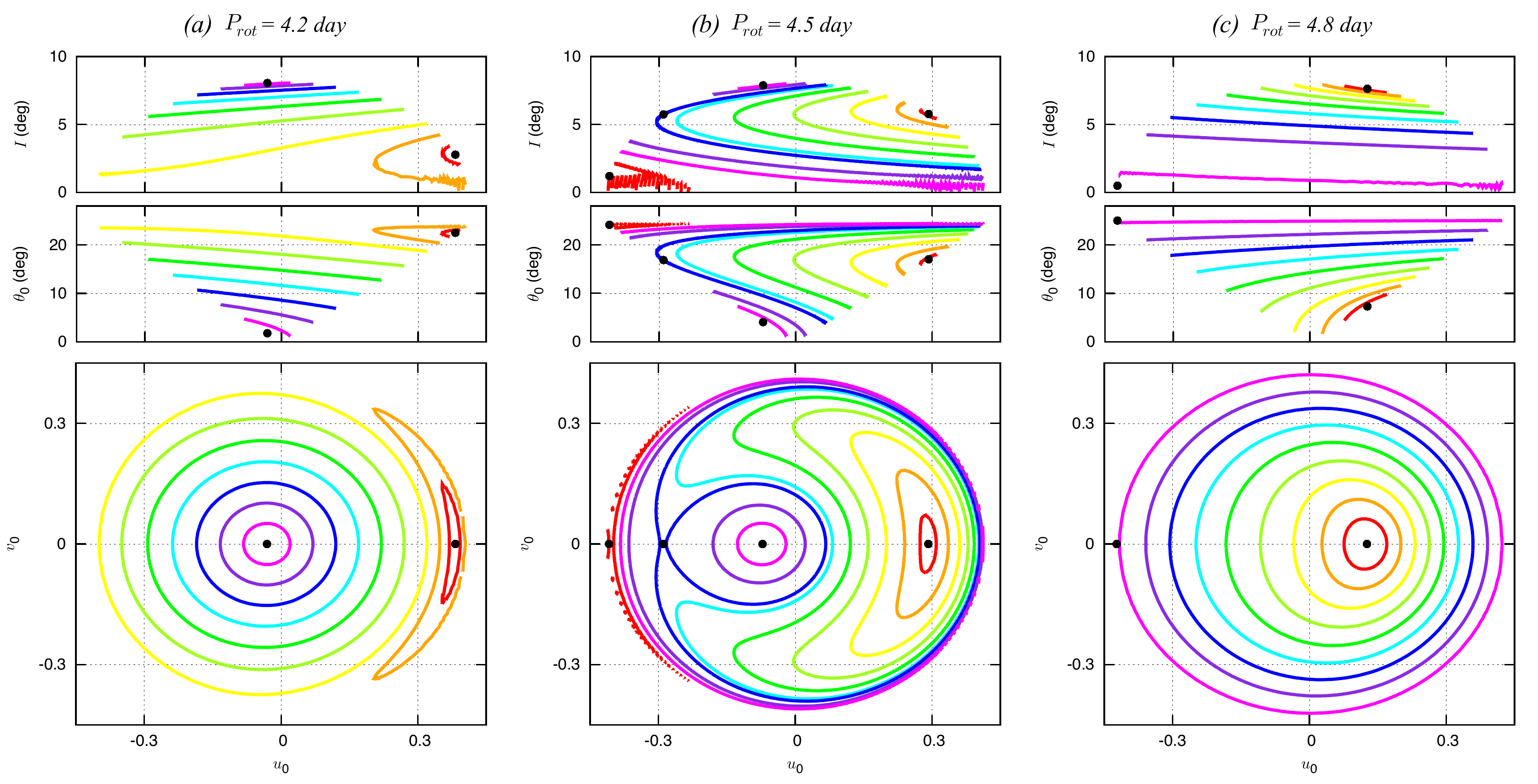} 
 \caption{Secular trajectories for the spin of the main star ($m_\p$) in the standard system (Table~\ref{standard}) for different rotation periods, (a) $P_\mathrm{rot} = 4.2$~day, (b) $P_\mathrm{rot} = 4.5$~day, and (c) $P_\mathrm{rot} = 4.8$~day.
We show the mutual inclination (top), the stellar spin projected on the orbit normal (middle), and its projection on the orbital plane (bottom). Cassini states are marked with a dot.
 \llabel{figSspin}  }
\end{sidewaysfigure}

In Figure~\ref{figSspin} we plot the level curves of the total energy in the plane $(\ui, \vi)$ for the standard system (Table\,\ref{standard}), adopting different rotation periods that range from 4.2 to 4.8~day.
Although these rotation periods may seem too short for Solar-type stars, they are reliable for young stars \citep[e.g.][]{Skumanich_1972}.
Moreover, close-binary systems undergo strong tidal effects that modify the rotation period until it is close to the orbital period, which corresponds to nearly 4.5~day in the case of the standard system (see section~\ref{tidalevol}).

For a rotation period of 4.8~day (Fig.\,\ref{figSspin}\,c), the stellar spin precesses around 
a direction close to the binary orbit's normal.
The obliquity is nearly constant as well as the mutual inclination between the two orbits.
However, for the transition rotation period of 4.5~day (Figs.\,\ref{figSspin}\,b), there is a resonance between 
the precession of the stellar spin and the precession of the orbits ($\alpha_\p \cos \de_\p / L_\p \approx \gamma_2 \cos I_0 / G_\a$), which completely modifies the evolution of the spin.
In this case, the obliquity is no longer constant, and the secular trajectories resemble those for the modified Kepler-16 system (Fig.\,\ref{figPspin}, right).
Moreover, unlike the Kepler-16's case, the mutual inclination between the orbits also undergoes significant variations.
Thus, for a star increasing its rotation rate, the spin can be captured in resonance (Fig.\,\ref{cassini}).
This event can produce a significant variations in the obliquity of the star and in the mutual inclination of the orbits.

\begin{figure}
\begin{center}
\includegraphics[width=\textwidth]{\figpath 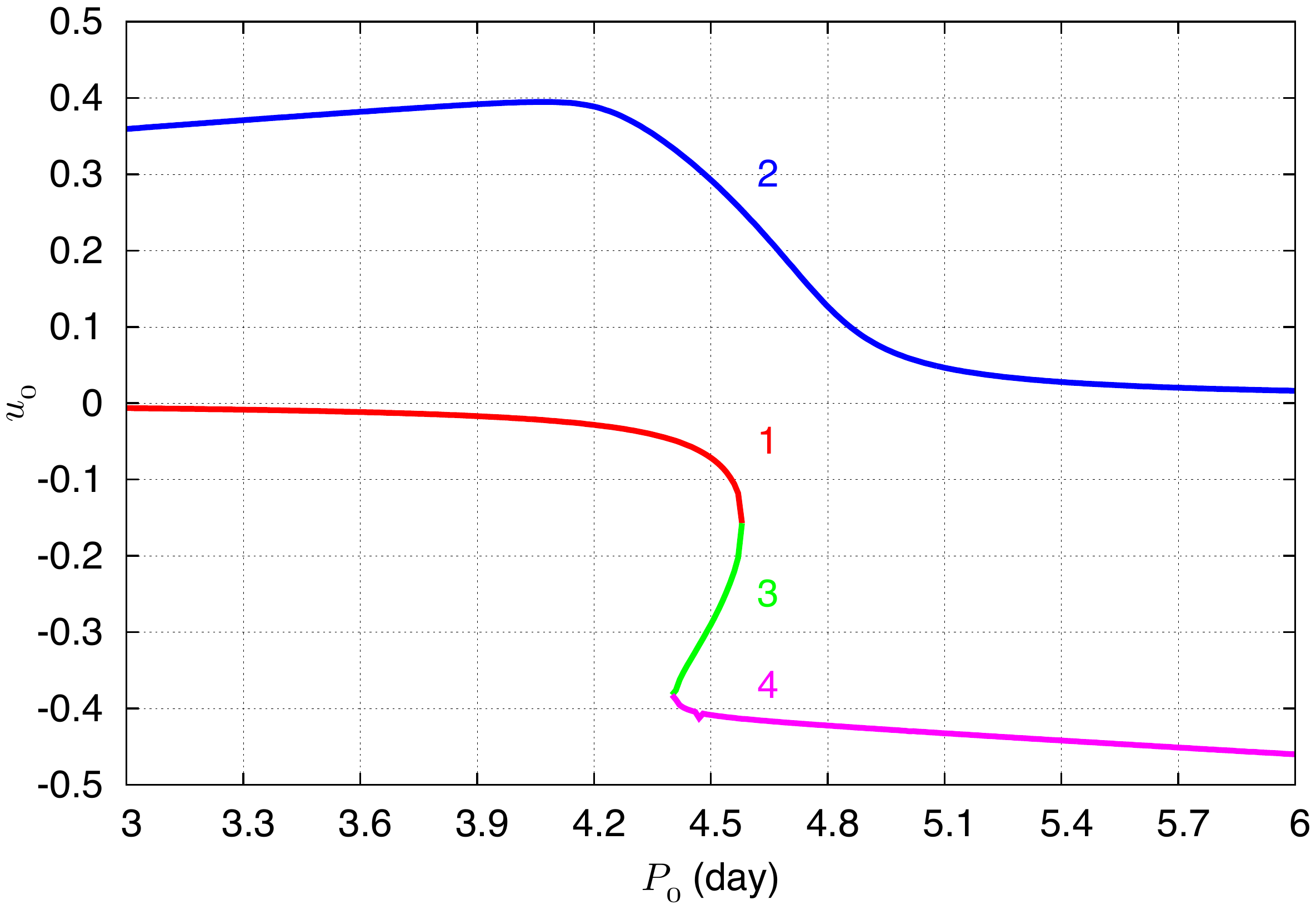} 
 \caption{Cassini states equilibria for the standard system (Table~\ref{standard}) as a function of the rotation period of the primary star, $P_\p$. These equilibria are obtained by solving equation (\ref{150514b}), and each color corresponds to a different dynamical state.
 \llabel{cassini}  }
\end{center}
\end{figure}

In Figure~\ref{cassini} we show the Cassini states equilibria for the standard system (Table~\ref{standard}) as a function of the rotation period of the primary star. 
This figure is equivalent to the classic Cassini states for the planet that are obtained  when solving equation (\ref{150513c}) as a function of the ratio $\alpha_\e / g_\e$ \citep[see, for example, Fig.~3 in][]{Ward_Hamilton_2004}.
We observe some differences in the number of Cassini states, but the key feature is the resonance near $P_0 \approx 4.5$~day, that allows two distinct evolutions for the stellar spin.

For close-in stars, in the expression of the total energy (\ref{150513d}), we also need to take into account the contribution of the secondary star: $ \moy{U_{R,\s}} = - \alpha_\s (\cos \de_\s)^2/2$.
In this case the problem is no longer integrable.
However, if the perturbation introduced by the secondary star dominates the perturbation from the planet ($ \alpha_\s \gg \gamma_2 $), we can neglect the term in $\gamma_2$ and the problem becomes integrable again \citep[see][]{Boue_Laskar_2009, Correia_2016}
by solving the following equations using the same steps explained in this section:
\be
U_R = \moy{U_{R,\p}} + \moy{U_{R,\s}}
 = - \frac{\alpha_\p}{2} (\cos \de_\p)^2 - \frac{\alpha_\s}{2} (\cos \de_\s)^2  \ ,
\llabel{150728x}
\ee
and
\be
L_\p G_\b \cos \de_\p + L_\s G_\b \cos \de_\s + L_\p L_\s \cos \de_{\p \s} = K_\p = cte \ ,
\llabel{150728w} 
\ee
where $\de_{\p \s}$ is the angle between the spin axis of the two stars \citep{Correia_2016}.



\section{Tidal evolution}
\llabel{tidalevol}

In this section we add the contribution of tides to the secular dynamics of circumbinary systems.
Tidal dissipation modifies the rotational and orbital angular momenta and the system evolves into some equilibrium configuration.
For the unknown geophysical parameters, we adopt for the stars $C/(m R^2) = 0.08$ and $k_2 = 0.028$ \citep{Eggleton_Kiseleva_2001}, and for the Jupiter-mass planets $C/(m R^2) = 0.25$ and $k_2 = 0.50$ \citep{Yoder_1995cnt}.
For tidal dissipation, we adopt the constant time-lag linear model presented in section~\ref{tidaleffects}.
For stars we use $\Delta t = 0.1$~s \citep{Penev_etal_2012}, and for planets $\Delta t = 100$~s \citep{Lainey_etal_2009}, which roughly corresponds to $Q \sim 10^7$ and $Q \sim 10^4$, respectively, with $ Q^{-1} \equiv n \Delta t$.

\subsection{Inner binary evolution}

We first look at the evolution of the inner binary without the presence of the companion planet, i.e., we restrict our analysis to the two body problem.
This simplification has been widely studied \citep[e.g.,][]{Kaula_1964, Goldreich_1966a, Alexander_1973, Efroimsky_Williams_2009, Correia_Laskar_2010B, Migaszewski_2012, Ferraz-Mello_2013, Correia_etal_2014, Makarov_2015}, but most previous studies focus on the spin and orbital evolution of a single component disturbed by a point-mass companion. 
This assumption is usually a good approximation while studying the evolution of a star-planet system.
However, it becomes less realistic for star-star systems, since both companions have similar angular momenta values.
As a consequence, in this case resonant interactions between the spin of the two stars can occur, which modify the intermediary evolution.

\begin{figure}
\begin{center}
\includegraphics[width=\textwidth]{\figpath 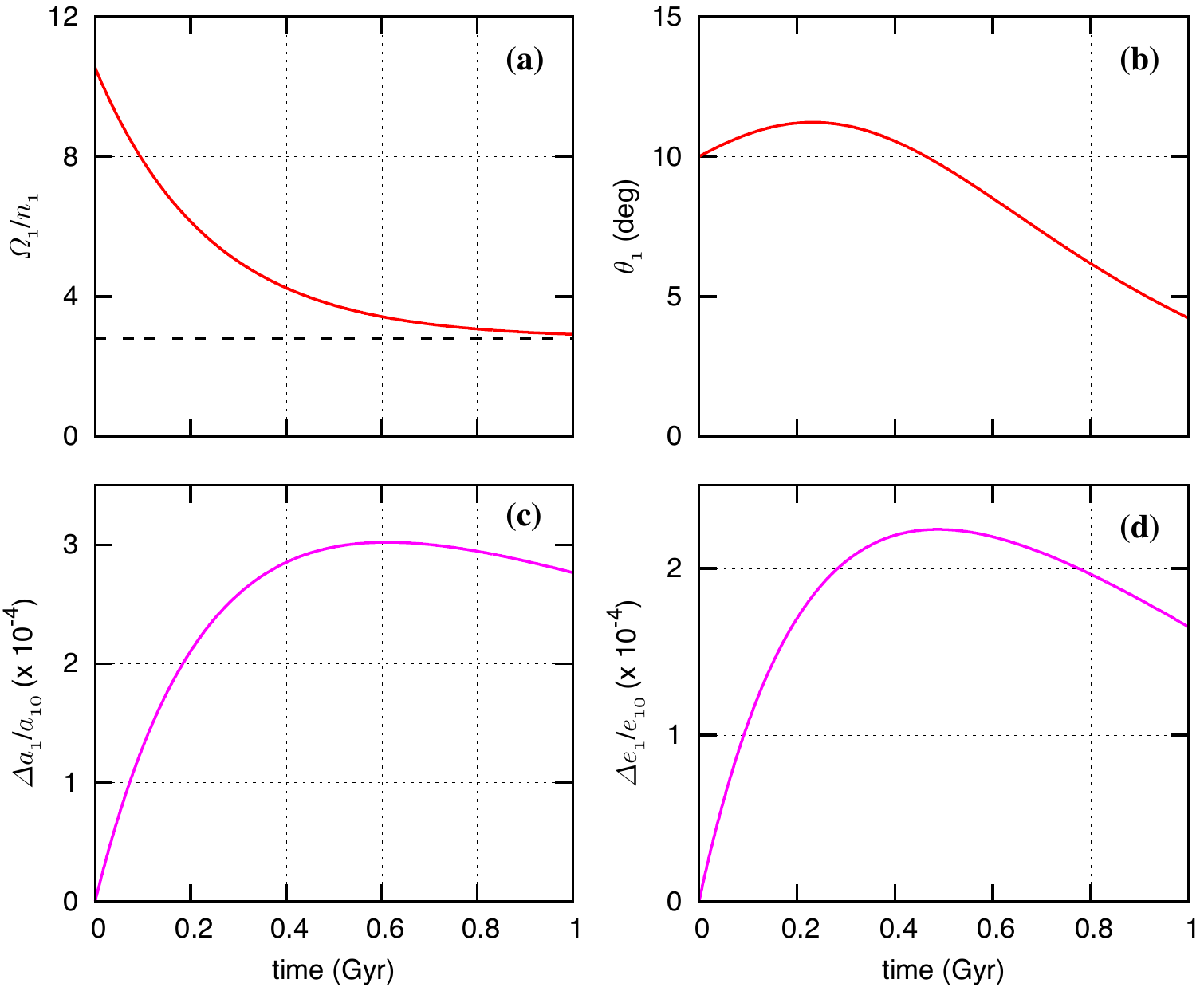} 
 \caption{Tidal evolution of the inner binary in the standard system (Table~\ref{standard}), for a point mass primary star with no spin ($k_{2,\p} = 0$), and in absence of the planetary companion ($m_\e = 0$). We show the relative rotation rate of the secondary star (a) and its obliquity (b), the relative semi-major axis variation (c), and the relative eccentricity variation (d). The dotted line corresponds to the equilibrium rotation given by expression (\ref{090520a}).
 \llabel{spintide}  }
\end{center}
\end{figure}

In Figure~\ref{spintide} we plot the tidal evolution of the inner binary in absence of the planetary companion for the standard system (Table~\ref{standard}).
The primary star is initially considered as a point mass with no spin, which is equivalent to assume $k_{2,\p} = 0$.
Therefore, all the modifications in the system result from tidal dissipation in the secondary star.
Its initial rotation period of is chosen to be 1~day, corresponding to $\om_\s/n_\b \approx 10.5$, while its initial obliquity is set at $\de_\s = 10^\circ$.
Tidal interactions with the primary decrease the rotation rate of the secondary (Fig.\,\ref{spintide}\,a) until it reaches the ``pseudo-synchronization'' equilibrium rotation \citep[e.g.,][]{Correia_2009}
\be
\frac{\om_\s^\eq}{n_\b} =  \frac{2 \cos \de_\s}{1 + \cos^2 \de_\s} \, \frac{f_2(e_\b)}{f_1(e_\b)} 
\ , \llabel{090520a}
\ee
while the obliquity tends to the equilibrium value
\be
\cos \de_\s  = \frac{2 n_\b}{\om_\s} \, \frac{f_2(e_\b)}{f_1(e_\b)}
\ . \llabel{090520dbis}
\ee
For initial fast rotation rates ($\om_\s \gg n_\b$), the equilibrium obliquity is close to $90^\circ$, that is why we observe an initial increase in the obliquity of the secondary star (Fig.\,\ref{spintide}\,b).
However, as the rotation rate slows down and approaches the equilibrium value (\ref{090520a}), the final obliquity tends to zero.

Concerning the orbit, the semi-major axis and eccentricity also initially increase, because the angular momentum is transferred from the spin to the orbit.
These variations are almost imperceptible, because in the example shown the orbital angular momentum is much larger than the rotational angular momentum of the secondary.
As the rotation rate comes close to the equilibrium value (\ref{090520a}), the angular momentum transfer ceases, and the only consequence of tidal dissipation is to decrease the orbital energy, hence we observe a decrease in the semi-major axis (Fig.\,\ref{spintide}\,c).
Since the orbital angular momentum must be conserved, i.e., $a_\b (1-e_\b^2) = cte. $ (Eq.\,\ref{130116c}), the eccentricity also decreases (Fig.\,\ref{spintide}\,d).
The final evolution of the system is obtained when the orbit becomes circular, the obliquity is zero, and the rotation rate is synchronous with the mean motion \citep[e.g.,][]{Hut_1980, Correia_2009}, although in most stellar systems this process takes longer than the maximum life-span of the stars.

\begin{figure}
\begin{center}
\includegraphics[width=\textwidth]{\figpath 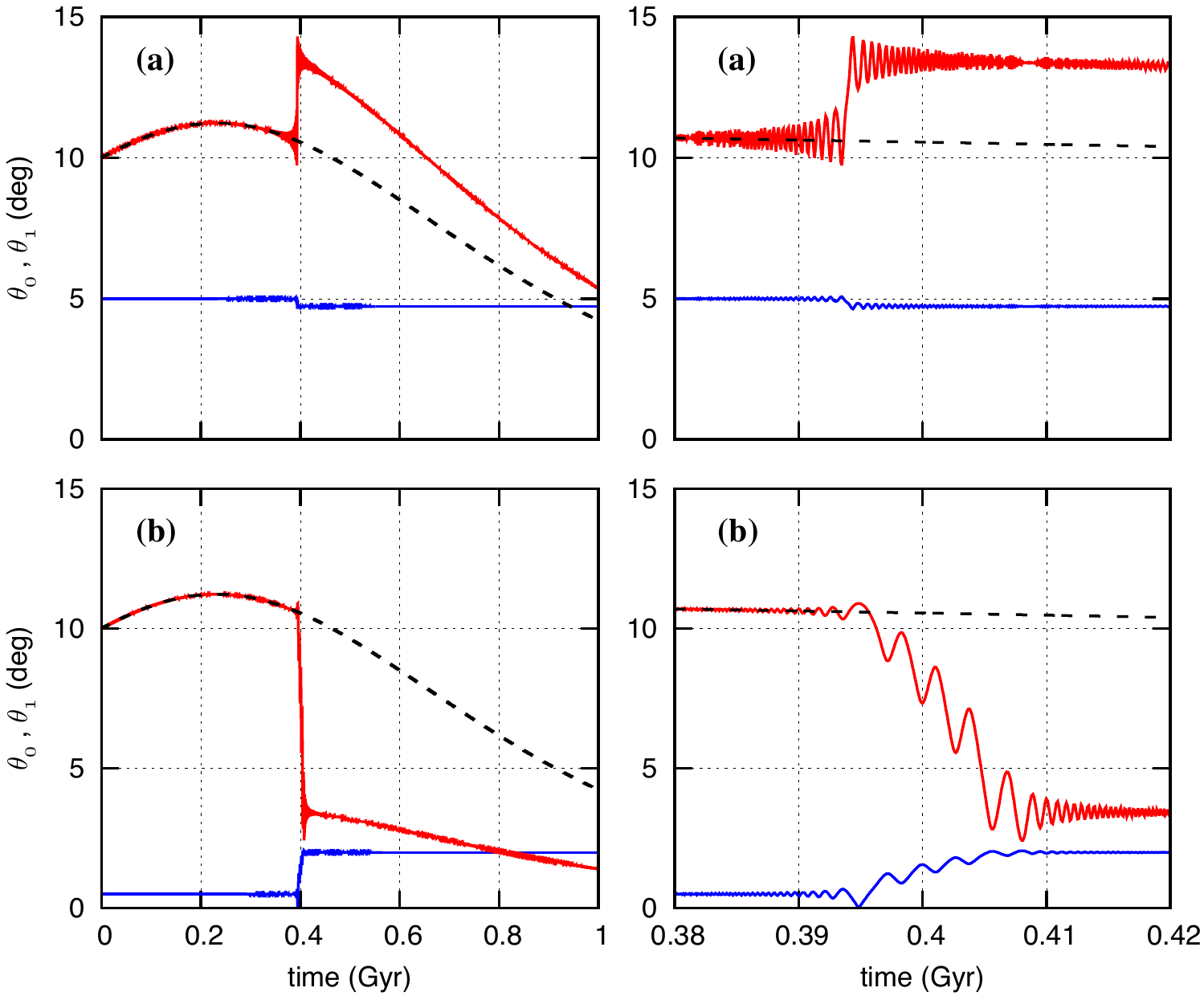} 
 \caption{Obliquity evolution of the inner binary in the standard system (Table~\ref{standard}), without dissipation in the primary star ($\Delta t_\p = 0$), and in absence of the planetary companion  ($m_\e = 0$).
We show the evolution for both stars using two different initial configurations of the primary star's spin: (a) initial obliquity $\de_\p = 5^\circ$; (b) initial obliquity $\de_\p = 0.5^\circ$. 
We plot the evolution through 1~Gyr (left), and near the resonance crossing around 0.4~Gyr (right).
The dotted line corresponds to the obliquity of the secondary for a point mass primary (Fig.\,\ref{spintide}\,b).
 \llabel{spinres}  }
\end{center}
\end{figure}

In Figure~\ref{spinres} we plot the evolution of the binary system when the primary star is no longer a point mass object.
The rotation of the primary is taken to be 10~day ($\om_\p \approx n_\b$), while for the initial obliquity we adopt $\de_\p = 5^\circ$ (Table~\ref{standard}), or 10 times smaller,  $\de_\p = 0.5^\circ$.
In order to better compare with the previous simulation, we neglect tidal dissipation on the primary star ($\Delta t_\p = 0$).
The orbital evolution and the evolution of the rotation rate are almost identical to those shown for a point mass companion (Fig.\,\ref{spintide}), so they are not shown.
However, the obliquity of the secondary can now experience quite different intermediary evolutions.
Indeed, around 0.4~Gyr, there is a resonance between the precession rates of the spin of both stars 
$  \alpha_\p \cos \de_\p / L_\p \approx \alpha_\s \cos \de_\s / L_\s$, which modifies their obliquities.
In one case, there is no capture in resonance, and the obliquities only receive a kick (Fig.\,\ref{spinres}\,a)
\citep[see also Appendix A in][]{Laskar_etal_2004E}.
In the other situation, capture occurs, and the obliquity of the secondary is brought near zero degrees in a time-scale much shorter than tidal effects alone (Fig.\,\ref{spinres}\,b).
The resonant equilibrium is only broken for low obliquity when the tidal torque becomes stronger than the precession torque.

After the resonant encounter, the obliquity of the secondary decreases again towards zero degrees.
The final evolution of the system is the same as the one described for a point-mass companion (Fig.\,\ref{spintide}), since it corresponds to the minimum of the total energy of the system \citep{Hut_1980}.
However, the time-scales involved can be significantly different.
Moreover, when additional bodies are present, such as a circumbinary planet, they will interact with a different stellar configuration that may drastically change their future evolution (see next section).
Therefore, when we inspect the tidal evolution of multi-body systems, we need to take their spin states into account.

\subsection{Planetary evolution}

The tides raised by the planet on the stars is much weaker than mutual tides between stars, so
we only take into account tidal effects raised by the stars on the planet (section~\ref{tidaleffects}).
These tides can be described as the tidal effect raised by the center of mass of the inner binary in the circumbinary planet (Eq.\,\ref{130122d}). 
Therefore, the tidal evolution of the planet is similar to the evolution of a single star described in previous section (see also Figure~\ref{spintide}).
Moreover, the orbital angular momentum of the circumbinary planet is much larger than the rotational angular momentum, so tides on the planet are only expected to significantly modify its spin.

In Figure~\ref{kepler16spin} we show the spin evolution of the planet Kepler-16\,b using the initial values from Table~\ref{standard}.
Although the orbits of this system are well established \citep{Doyle_etal_2011}, we ignore the present values for the rotation period and obliquity.
Assuming initial fast rotating planets like the gaseous planets of the solar system, the general trend for the spin evolution corresponds to a progressive increase in the rotation period and obliquity as the one shown in Figure~\ref{kepler16spin}.
In the particular case of Kepler-16\,b, we additionally observe a small kick on its obliquity around $t \approx 0.6$~Gyr. 
This corresponds to a secular resonance between the precession of the spin and the frequency $ p -g_1+g_2 \approx - 3.58'$/yr, where $p$ is the precession frequency of the node, and $g_1$, $g_2$ are the precession frequencies for the pericenters of both orbits.

\begin{figure}
\begin{center}
\includegraphics[width=\textwidth]{\figpath 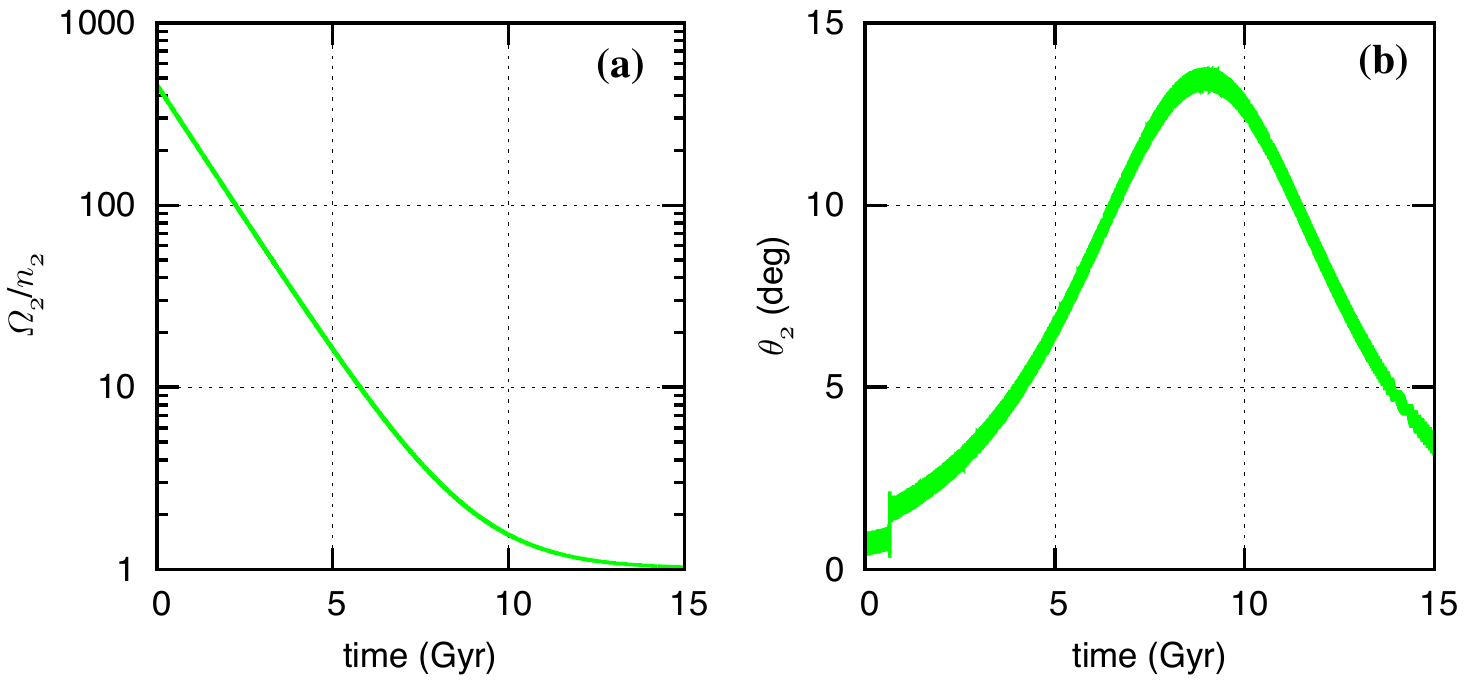} 
 \caption{Tidal evolution of the planet Kepler-16\,b (Table~\ref{standard}). We show the relative rotation rate (a), and the obliquity (b).
\llabel{kepler16spin}  }
\end{center}
\end{figure}

\subsection{Coupled evolution}

For binary systems with separations $a_\b > 0.1$~AU and moderate eccentricities, the orbits are only marginally modified by tides during the age of the system (Fig.~\ref{spintide}c,d).
The spins of all bodies can be significantly modified, but they present a general trend of synchronising the rotation with the orbit and decreasing the obliquity (Fig.~\ref{spintide}a,b).
However, the spin can cross some secular resonances that may accelerate or delay its evolution (Figs.~\ref{spinres} and~\ref{kepler16spin}).
In the examples previously shown, these resonances had no impact on the orbits, either because they corresponded to spin-spin interactions (Fig.~\ref{spinres}), or because the orbital angular momentum is much larger than the rotational angular momentum (Fig.~\ref{kepler16spin}).

\subsubsection{Secular spin-orbit resonances}

In Figure~\ref{standtide} we show the tidal evolution of the standard system (Table~\ref{standard}).
Unlike the previous examples, we observe some resonant interactions that considerably modify the spins and the orbits.
The main interactions correspond to 1) a resonance between the spin precession rate of the secondary and the precession of the node at $t \approx 85$~Myr; 2) a resonance between the spin precession rate of the primary and the precession of the node for $ 150 \lesssim t \lesssim 250 $~Myr.
A spin-spin resonance between the spin precession rates of both stars at $t \approx 125$~Myr is also noticeable.

\begin{figure}
\begin{center}
\includegraphics[width=\textwidth]{\figpath 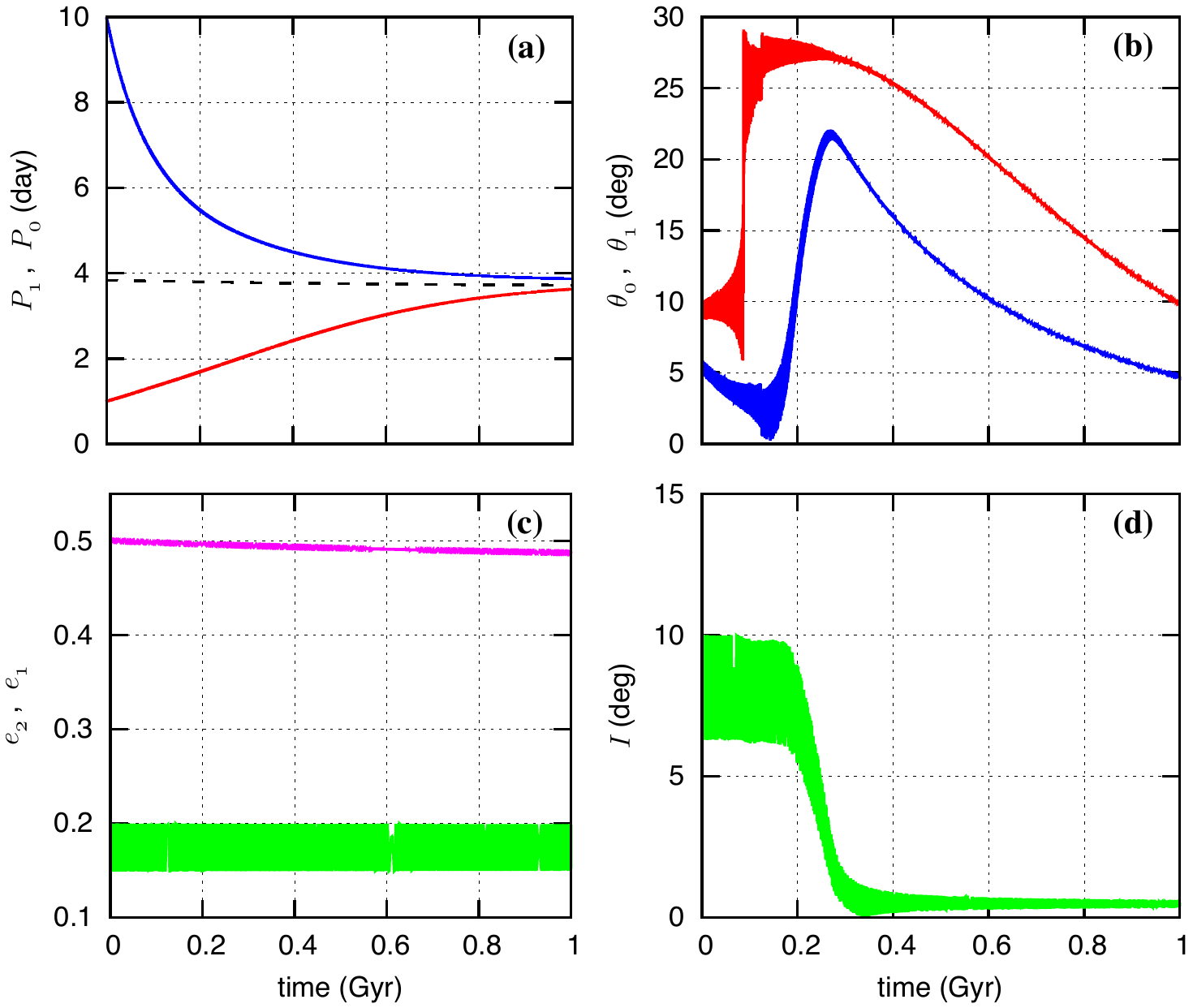} 
 \caption{Tidal evolution of the standard system (Table~\ref{standard}). We show the rotation rate of the stars (a), their obliquity (b), the eccentricity of the orbits (c), and the mutual inclination (d). The dotted line corresponds to the equilibrium rotation given by expression (\ref{090520a}).
\llabel{standtide}  }
\end{center}
\end{figure}

The two spin-orbit resonances result in an obliquity increase of about 20 degrees. 
However, while for the secondary the increase occurs in a short time-scale, less than 1~Myr, for the primary it lasts for more than 100~Myr.
The different behaviors correspond to two different kinds of resonance crossing.
For the secondary, the rotation period is increasing from lower values, while for the primary it is the opposite (Fig.~\ref{standtide}a).
Therefore, the secondary cannot be trapped in resonance, while for the primary this is possible (Fig.~\ref{cassini}).
The two different evolutions can be well understood using the diagrams with all possible secular trajectories given in Figure~\ref{figSspin}.
These diagrams correspond to the spin of the primary, but for the secondary we obtain a similar picture, only the rotation period for which the resonance is crossed would change ($\approx1.5$~day, instead of 4.5~day).

\begin{figure}
\begin{center}
\includegraphics[width=\textwidth]{\figpath 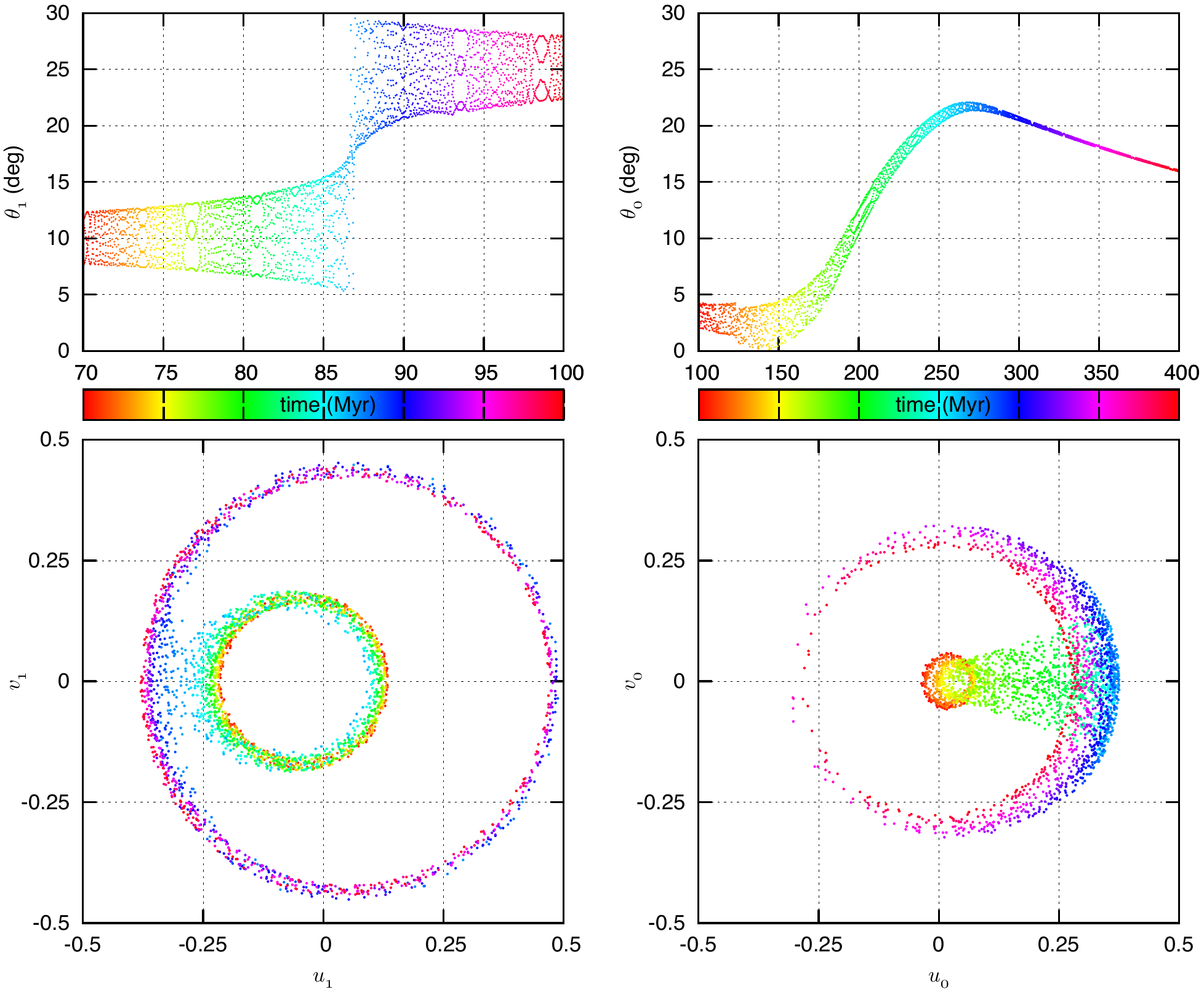} 
 \caption{Tidal evolution of the standard system (Table~\ref{standard}). We show the obliquity (top) and the projection of the spin on the orbital plane (bottom) of the secondary (left) and of the primary (right)  during the resonance crossing. Each dot corresponds to a different time. 
\llabel{cassTide}  }
\end{center}
\end{figure}

In Figure~\ref{cassTide} we show the projection of the spin of both stars during the resonant encounter.
For the secondary, the spin is initially circulating around state~1.
As the the rotation period increases, this state merges with the hyperbolic state~3 and the resonant equilibria disappear.
The spin is then forced to circulate around state~2, with higher obliquity (Fig.~\ref{cassini}).

For the primary, the spin is initially circulating around state~2.
As the rotation period decreases, it crosses the resonance around $P_\p \approx 4.6$~day.
At this point, the spin can either start circulating around state~1 or follow librating around state~2, as it happens in our example.
For rotation periods faster than the resonant period, the obliquity of state~2 increases (Fig.~\ref{cassini}).
As long as tidal dissipation is not too strong, the evolution is adiabatic and the spin remains trapped in resonance (state 2).
Therefore, although tidal effects tend to damp the obliquity, we observe that the obliquity actually increases such that the resonant ratio can be maintained.

More interestingly, we also observe that while the obliquity of the primary increases, the mutual inclination is damped.
This unexpected behavior is also a result of a change in the topology of the system.
Indeed, in Figure~\ref{figSspin} we can see that while the obliquity of state 2 increases for faster rotation periods, the corresponding mutual inclination decreases.
The equilibrium inclination for a given Cassini state can be obtained directly from expression (\ref{141217f}).
In the case of the standard system we have $L_\p \ll G_\a \ll G_\b$, so we can simplify this expression as
\be
\cos I \approx  \Z \approx \frac{K_\p}{G_\b G_\a} - \frac{L_\p}{G_\a}  \cos \de_\p  \ .
\llabel{151207a} 
\ee
Assuming $L_\p$ approximately constant, we see that an increase in the obliquity corresponds to a decrease of the mutual inclination and vice versa.
However, $L_\p$ is not completely constant, since the rotation period is varying during the resonant crossing.
Therefore, in Figure~\ref{cassinc} we plot the exact solution of equation (\ref{141217f}) as a function of the rotation period, where the obliquity curve corresponds to the obliquity of state~2 shown in Figure~\ref{cassini}.
We observe there is a very good agreement with the numerical simulations shown in Figure~\ref{standtide}.
The small differences observed result from the fact that our analytical model is obtained with the quadrupolar approximation for the orbits and considering only the spin of the primary.

\begin{figure}
\begin{center}
\includegraphics[width=\textwidth]{\figpath 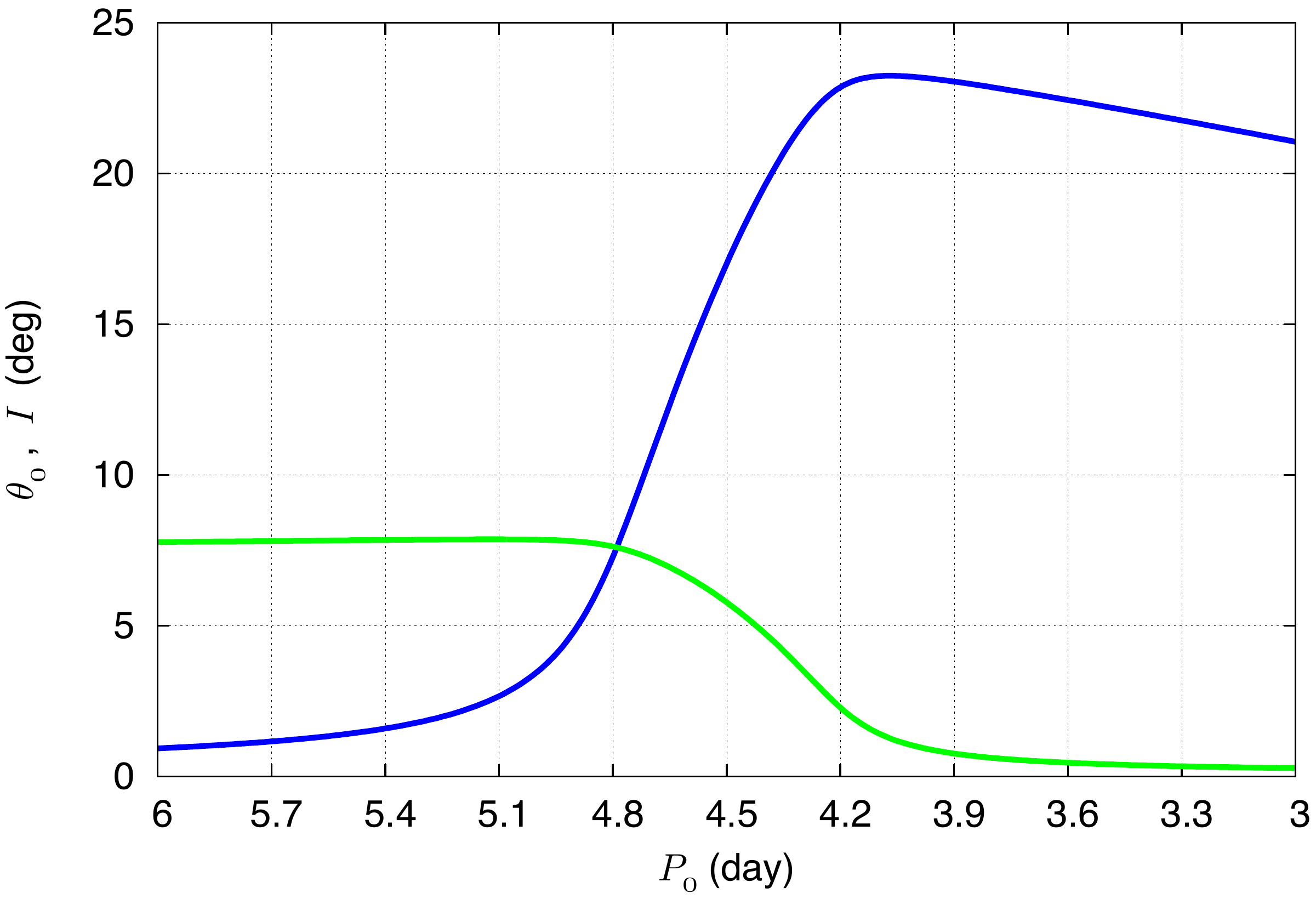} 
 \caption{Cassini state~2 equilibria for the standard system (Table~\ref{standard}) as a function of the rotation period of the primary star, $P_\p$. These equilibria are obtained by solving equation (\ref{150514b}). We show the obliquity of the primary (blue) and the mutual inclination (green).
\llabel{cassinc}  }
\end{center}
\end{figure}

As the rotation period decreases, the libration width of the resonant state~2 decreases (Fig.\,\ref{figSspin}a).
When the obliquity is near to its maximum value (and the mutual inclination is near zero), the tidal torque becomes stronger than the resonant coupling.
At this stage, the spin ends its libration around state~2, and starts to circulate around state~1.
In absence of the secular resonant forcing, the obliquity is damped by tides until it reaches the equilibrium Cassini state~1 value.

In Figure~\ref{standtideinc} we show the evolution of the obliquity and the mutual inclination for a modified standard system with an initial inclination of $60^\circ$ (instead of $10^\circ$).
Although this new configuration presents larger oscillations of the mutual inclination (due to the secular quadrupolar interactions between the two orbits), its average value is also damped to zero while crossing the secular resonance with the spin of the primary.
For larger initial mutual inclinations we would observe a similar behavior, at least as long as the pericenter of the inner orbit is in circulation (Fig.\,\ref{figquadi}).
The damping of the mutual inclination is a consequence of the angular momentum transfer between the spin of the primary and the orbit of the planet through the secular resonance.
We thus conclude that this resonance is an efficient mechanism of transforming circumbinary systems with arbitrary initial mutual inclination in coplanar systems like the ones observed by the {\it Kepler Spacecraft} surveys \citep[e.g.][]{Doyle_etal_2011, Welsh_etal_2012, Orosz_etal_2012a, Orosz_etal_2012b, Kostov_etal_2016}.

\begin{figure}
\begin{center}
\includegraphics[width=\textwidth]{\figpath 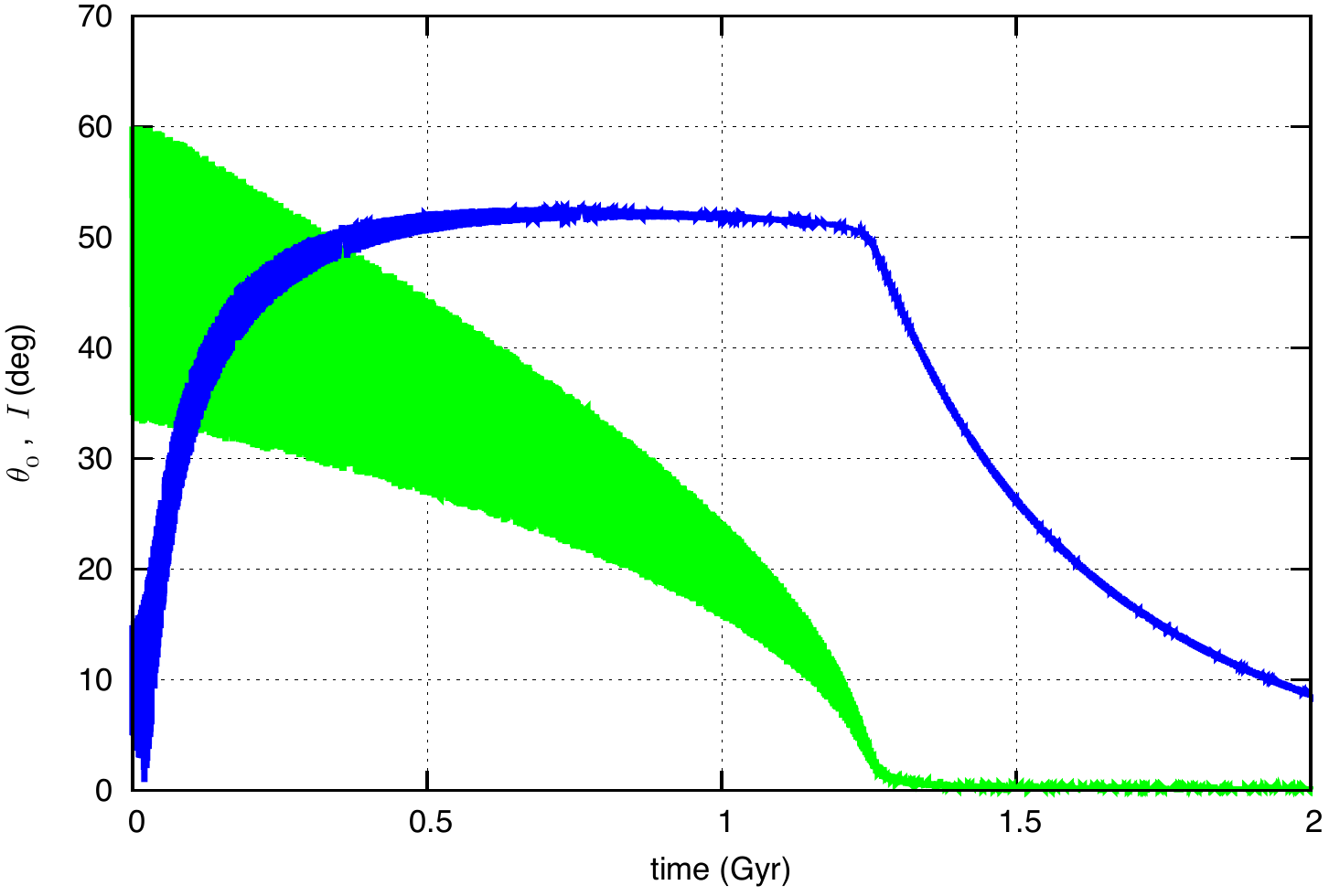} 
 \caption{Tidal evolution of the standard system (Table~\ref{standard}) with initial mutual inclination $I=60^\circ$. We show the obliquity of the primary (blue) and the mutual inclination (green).
\llabel{standtideinc}  }
\end{center}
\end{figure}

\subsubsection{Chaotic evolution}

In section~\ref{orbdyn} we saw that there are two different dynamical regimes for the orbits: the pericenter of the inner orbit, $\w_\b$, can be in circulation or in libration around $\pm 90^\circ$ (Fig.\,\ref{figquadi}).
In previous section we have increased the initial mutual inclination of the standard system up to $60^\circ$, but we kept $\w_\b = 0^\circ$ (Fig.\,\ref{standtideinc}).
As a consequence, the orbits were always in circulation (Fig.\,\ref{figquadi}).

We now keep initial $I=60^\circ$, but with initial $\w_\b=90^\circ$ for the standard system (Table~\ref{standard}), which places the system in libration.
In Figure~\ref{standlib} we show two different evolutions with slightly different initial precession angle for the primary star, $\phi_\p=0.0^\circ$ (a) and $\phi_\p=0.1^\circ$ (b).

\begin{figure}
\begin{center}
\includegraphics[width=\textwidth]{\figpath 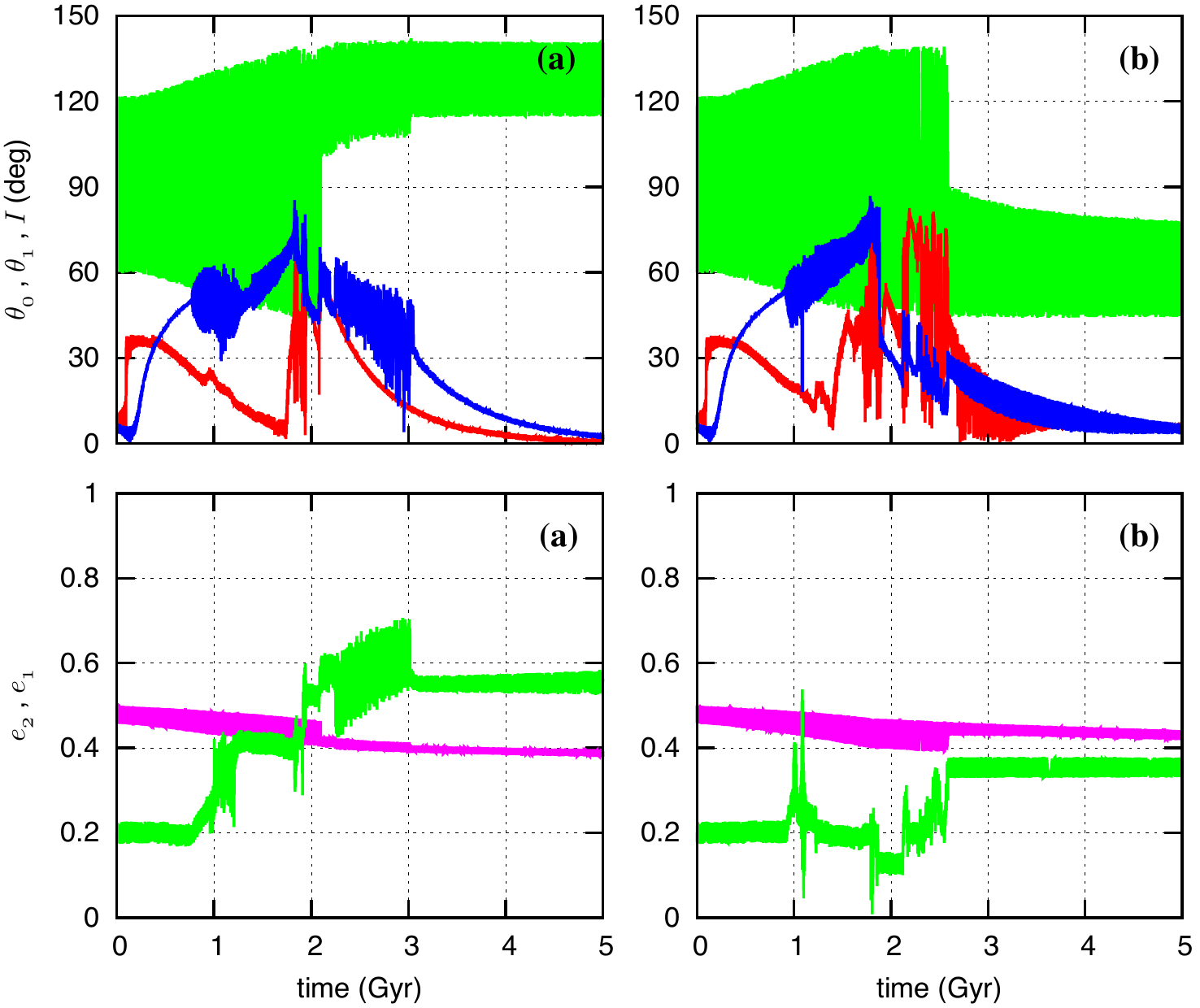} 
 \caption{Tidal evolution of the standard system (Table~\ref{standard}) with initial $I=60^\circ$, $\w_\b=90^\circ$, and $\phi_\p=0.0^\circ$ (a) or $\phi_\p=0.1^\circ$ (b).
We show the obliquity of the stars together with the mutual inclination (top), and the eccentricity of the orbits (bottom).
\llabel{standlib}  }
\end{center}
\end{figure}

The initial evolution of the system is similar to the circulation regime (Fig.\,\ref{standtide}): the average mutual inclination is constant, the spin of the secondary encounters a resonance at $t \approx 0.1$~Gyr, and the primary at $t \approx 0.2$~Gyr.
At this stage, the obliquity of the primary increases, which is accompanied by an increase in the amplitude of the mutual inclination.
The mechanism behind this exchange is the same described for the circulation regime (see Fig.\,\ref{cassinc}).
As the amplitude of the mutual inclination increase, the system approaches the separatrix between the libration and circulation regimes  (Fig.\,\ref{figquadi}).
Around $t \approx 1$~Gyr the stellar spins and the eccentricity of the planet's orbit become excited by several resonances and become chaotic.

The chaotic regime can last for several Gyr, until the separatrix is crossed and the system enters in circulation.
For long times spent in the libration region near the separatrix, the eccentricity of the outer orbit can rise to values close to the unity, so the planet can experiment close encounters with the remaining bodies and the system becomes unstable. 
In this case the secular model presented in this paper is no longer adapted to follow the orbits.
However, as soon as the system enters in the circulation regime, the eccentricity of the planet's orbit stabilises, as well as the spins.
The obliquity of all bodies is strongly chaotic in the libration region, but it becomes dominated by tides in the circulation region.

The evolution in the chaotic zone is very sensitive to the initial conditions. 
In the example shown in Figure~\ref{standlib} we have just modified slightly the initial precession angle of the spin of the primary by $\Delta \phi_\p = 0.1^\circ$.
We have run many more simulations, with different phase angles, and also with faster and slower initial rotation rates for the stars.
In half of the cases, the separatrix is crossed with inclination larger than $90^\circ$, so the orbit of the planet becomes retrograde (Fig.\,\ref{standlib}\,a), while in another half the system becomes prograde (Fig.\,\ref{standlib}\,b).
The initial resonant crossing at $t < 0.2$~Gyr can be avoided, but we observed that for systems initially in libration, the spins always become excited after some time.
As a consequence, in all runs the amplitude of the mutual inclination increased and the systems ultimately quit the libration regime.
Their final evolution is therefore always similar to the one shown in Figure~\ref{standlib}. 
We hence conclude that tidally evolved circumbinary systems are most likely in circulation.

In Figure~\ref{standsep} we show two evolutions for the standard system (Table~\ref{standard}) starting with initial $I=43.5^\circ$ and $\w_\b=90^\circ$.
These initial conditions still correspond to the circulation regime, but place the system very close to the separatrix.
We adopt again slightly different initial precession angle for the primary star, $\phi_\p=0.0^\circ$ (a) and $\phi_\p=0.1^\circ$ (b). 
Moreover, we adopt an initial fast rotation period for the primary, $P_\p=1.5$~day. 
Thus, tidal effects will increase the semi-major axis and the eccentricity of the inner orbit during the initial stages (Fig.\,\ref{spintide}), which forces the system to cross the separatrix and enter in the libration region.

\begin{figure}
\begin{center}
\includegraphics[width=\textwidth]{\figpath 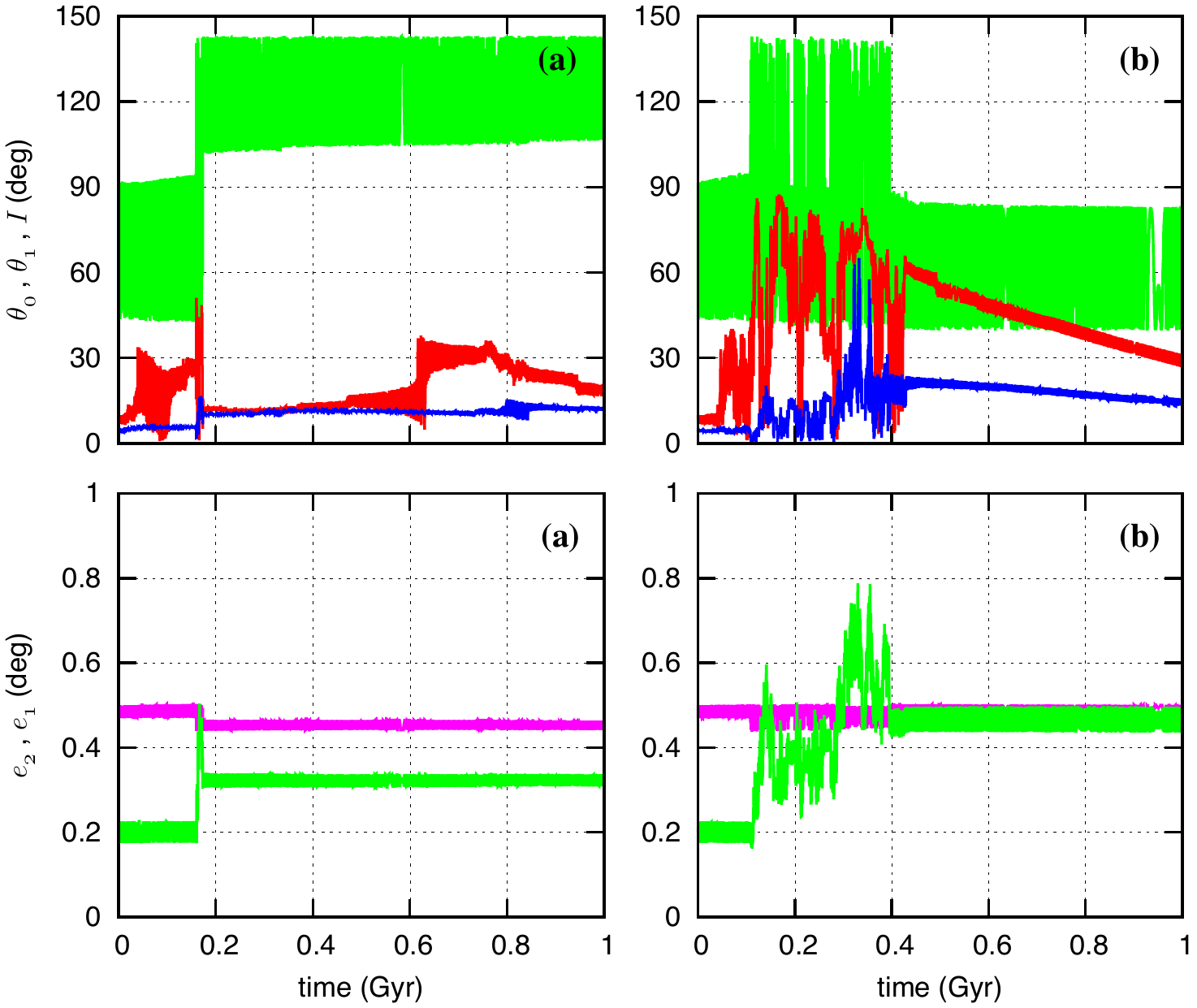} 
 \caption{Tidal evolution of the standard system (Table~\ref{standard}) with initial $I=43.5^\circ$, $\w_\b=90^\circ$, $P_\p=1.5$~day,
and $\phi_\p=0.0^\circ$ (a) or $\phi_\p=0.1^\circ$ (b).
We show the obliquity of the stars together with the mutual inclination (top), and the eccentricity of the orbits (bottom).
\llabel{standsep}  }
\end{center}
\end{figure}

Once the system enters in libration, the mutual inclination undergoes variations ranging from $44^\circ$ to $136^\circ$ (Fig.\,\ref{figquadi}), meaning that the orbit of the planet oscillates from prograde to retrograde.
As in the previous example (Fig.\,\ref{standlib}), the spins of all bodies and the eccentricity of the outer orbit become chaotic.
As the rotation period of the stars approach their equilibrium values, the semi-major axis and the eccentricity of the inner orbit decrease again and the system returns to circulation.
However, in half of the cases, the separatrix can be crossed with high inclination and the orbit of the planet becomes retrograde (Fig.\,\ref{standsep}\,a).
This mechanism is then able to transform previous prograde orbits in retrograde ones, as it was already reported by \citet{Correia_etal_2011}.


\section{Additional applications}
\llabel{otherappli}

The secular model presented in this paper (section~\ref{secmodel}) is very general and therefore its validity is not restricted to circumbinary systems.
Indeed, it can also be applied to any three-body hierarchical system for which terms in $(r_\b/r_\a)^4$ can be neglected (octupolar approximation), as well as the torque of the outermost body on the spin of the inner bodies.
Nevertheless, this last effect can also be considered as in \citet{Correia_etal_2011}, by keeping in the rotational potential (\ref{130116a}) the contribution from $\M$ (Eq.\,\ref{130118y}):
\be
U_{R,\ij} = G \frac{m_\p m_\s}{r_\b} J_{2,\ij} \left(\frac{R_\ij}{r_\b}\right)^2 \left[ P_2 (\vv{\hat r}_\b \cdot \vv{\hat \k}_\ij) + \frac{m_\e}{m_\jk} \left(\frac{r_\b}{r_\a} \right)^3 P_2 (\vv{\hat r}_\a \cdot \vv{\hat \k}_\ij) \right]  \ .
\llabel{151217a}
\ee

A straightforward application is, for instance, the formation of hot Jupiters from secular planet-planet interactions \citep{Naoz_etal_2011, Beauge_Nesvorny_2012}.
In Figure~\ref{naoz} we reproduce a simulation for a Sun-like star with two Jupiter-mass planetary companions using the same initial conditions as for Figure~2 by \citet{Naoz_etal_2011}.
We observe there is a good agreement between the two models\footnote{In order to reproduce the results in \citet{Naoz_etal_2011} we cannot take into account the flattening of the star (Eq.\,\ref{130116a}). The evolution is also highly chaotic, so a slightly change in the reference angles lead to different final mutual inclination.}.

Another suitable applications are star-planet-satellite systems and triple stellar systems.

\begin{figure}
\begin{center}
\includegraphics[width=\textwidth]{\figpath 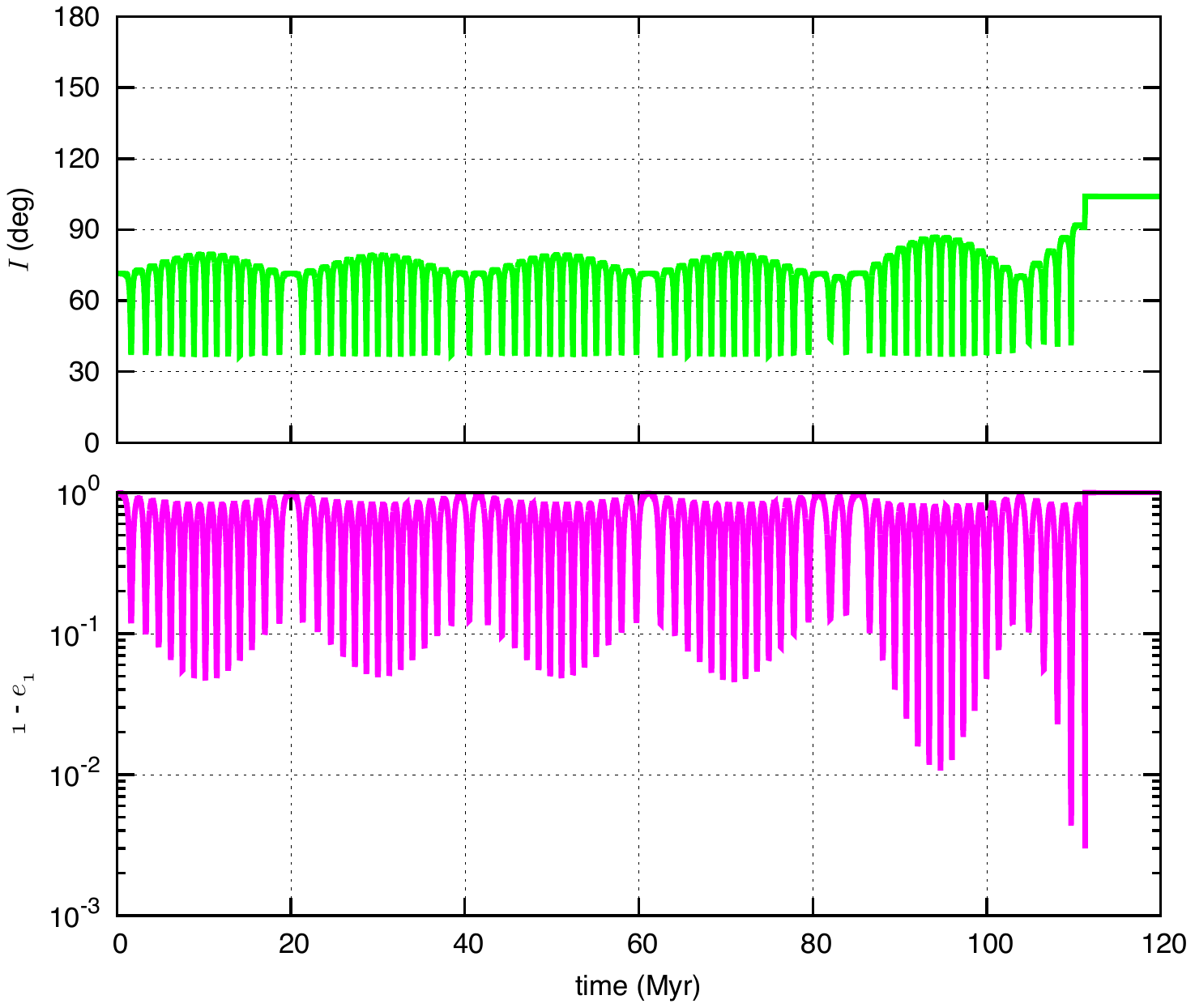} 
 \caption{Tidal evolution of a fictitious system with a Sun-like star and two Jupiter-mass planetary companions using the same initial conditions as for Figure~2 in \citet{Naoz_etal_2011}.
\llabel{naoz}  }
\end{center}
\end{figure}

\section{Conclusion}

In this paper we provide a secular model to describe the evolution of circumbinary three-body systems, where all bodies undergo tidal interactions.
We use the octupolar non-restricted approximation for the orbital interactions, including general relativity corrections, the quadrupolar approximation for the spins, and the viscous linear model for tides. 

Many of the already known circumbinary planets evolve in coplanar orbits, since they were detected through the transits method.
However, this is not necessarily the standard configuration for circumbinary planets \citep[e.g.][]{Martin_Triaud_2014}.
Our model is suitable to study the long-term evolution of a wide variety of circumbinary systems in very eccentric and inclined orbits.

We have shown that tidal effects coupled with the secular evolution can totally modify the final configuration of the system.
For instance, tides alone are unable to damp the mutual inclination of planetary systems during their life-times.
A most striking example is that circumbinary planets with initial arbitrary orbital inclination can become coplanar through a secular resonance between the precession of the spin of one star and the precession of the orbit.
We also show that circumbinary systems for which the pericenter of the inner orbit is initially in libration present chaotic motion for the spins and for the eccentricity of the outer orbit.

We have presented in this paper a few examples, which are representative of the diversity of behaviors among circubinary systems.
Many other systems are awaiting to be studied.
The fact that we use average equations for both tidal and gravitational effects, makes our method suitable to be applied in long-term studies.
It allows to run many simulations for different initial conditions in order to explore the entire phase space and evolutionary scenarios.
In particular, it can be very useful to derive constraints for the past and future tidal evolution of circumbinary systems.
Our model can also be applied to any three-body hierarchical system such as star-planet-satellite systems and triple stellar systems.

\begin{acknowledgements}
We acknowledge support from PNP-CNRS, and from from CIDMA strategic project UID/MAT/04106/2013.
\end{acknowledgements}

\appendix

\section{Oblate spheroid potential}

\llabel{apenB}

The gravitational potential of an oblate body of mass $m_\ij$ symmetric about its rotation axis $\vv{\hat \k}$ is given by \citep[e.g.][]{Goldstein_1950}:
\be
V_{\ij} (\vv{r}) = - G \frac{m_\ij}{r} \left[ 1 - J_{2,\ij} \left(\frac{R_\ij}{r}\right)^2 P_2 (\vv{\hat r} \cdot \vv{\hat \k}_\ij) \right]
\ , \llabel{130118a}
\ee
where we neglected terms in $(R_\ij/r)^3$.  
The gravity field coefficient $J_{2,\ij}$ is obtained from the principal moments of inertia $A_\ij = B_\ij$ and $C_\ij$ as $J_{2,\ij} = (C_\ij-A_\ij) / m_\ij R_\ij^2$.
When the asymmetry in the body mass distribution results only from its rotation,  $J_{2,\ij}$
is given by expression (\ref{101220a}).
The main term in the above expression is responsible for the orbital motion (Eq.\,\ref{090514a}), while the contribution in $J_{2,\ij}$ is responsible for a perturbation of this motion, since $ J_{2,\ij} (R_\ij/r)^2 \ll 1 $.
Thus, retaining only the terms in $J_{2,\ij}$, the resulting perturbing potential energy of a system composed of three oblate bodies is given by:
\be
U_R = U_{R,\p} + U_{R,\s} + U_{R,\e}
\ , \llabel{130118b}
\ee
where we have for the planet
\be
U_{R,\e} = \sum_{\ij = 0,1} m_\ij V_{\e} (\vv{r}_{\a \ij}) =  \sum_{\ij = 0,1} G \frac{m_\ij m_\e}{r_{\a \ij}} J_{2,\e} \left(\frac{R_\e}{r_{\a \ij}}\right)^2 P_2 (\vv{\hat r}_{\a \ij} \cdot \vv{\hat \k}_\e) 
\ , \llabel{130118d}
\ee
and for each star ($\ij = 0, 1$)
\be
U_{R,\ij} =  G \frac{m_\p m_\s}{r_\b} J_{2,\ij} \left(\frac{R_\ij}{r_\b}\right)^2 P_2 (\vv{\hat r}_\b \cdot \vv{\hat \k}_\ij) 
+ G \frac{m_\ij m_\e}{r_{\a \ij}} J_{2,\ij} \left(\frac{R_\ij}{r_{\a \ij}}\right)^2 P_2 (\vv{\hat r}_{\a \ij} \cdot \vv{\hat \k}_\ij)
\ . \llabel{130118c}
\ee
We also have (Fig.\,\ref{fig1})

\be
 \vv{r}_{\a \ij} = \vv{r}_\a + \delta_\ij \vv{r}_\b
\ , \llabel{130118e}
\ee
where $ \delta_\p = m_\s / \m $ and $ \delta_\s = - m_\p / \m $.
Since we assume that $r_\b \ll r_\a$, we can write
\be
\frac{P_2 (\vv{\hat r}_{\a \ij} \cdot \vv{\hat \k}_\ji)}{r_{\a \ij}^3} \approx
\frac{P_2 (\vv{\hat r}_\a \cdot \vv{\hat \k}_\ji)}{r_\a^3} 
+ \frac{3 \delta_\ij}{2 r_\a^3} \frac{r_\b}{r_\a} \left[ 
\vv{\hat r}_\b \cdot \vv{\hat r}_\a
- 5 (\vv{\hat r}_\b \cdot \vv{\hat r}_\a) (\vv{\hat r}_\a \cdot \vv{\hat \k}_\ji)^2
+ 2 (\vv{\hat r}_\b \cdot \vv{s}_\ji) (\vv{\hat r}_\a \cdot \vv{\hat \k}_\ji)
\right] \ , \llabel{130118f}
\ee
where we neglected terms in $(r_\b / r_\a)^2$, that is, we neglect terms in $J_{2,\ij} (R_\ij/r_\ij)^2 (r_\b / r_\a)^2$ in the potential energy.
Replacing in expressions (\ref{130118c}) and (\ref{130118d}) we get for the planet
\be
U_{R,\e} =  G \frac{m_\e \m}{r_\a} J_{2,\e} \left(\frac{R_\e}{r_\a} \right)^2 P_2 (\vv{\hat r}_\a \cdot \vv{\hat \k}_\e) 
\ , \llabel{130118z}
\ee
since 
$m_\p \delta_\p + m_\s \delta_\s = 0$, and for each star ($\ij = 0,1$; $\jk=1-\ij$)

\begin{eqnarray}
U_{R,\ij}  & =&  
G \frac{m_\p m_\s}{r_\b} J_{2,\ij} \left(\frac{R_\ij}{r_\b}\right)^2 \left[ P_2 (\vv{\hat r}_\b \cdot \vv{\hat \k}_\ij) + \frac{m_\e}{m_\jk} \left(\frac{r_\b}{r_\a} \right)^3 P_2 (\vv{\hat r}_\a \cdot \vv{\hat \k}_\ij) \right] \crm &\approx&
G \frac{m_\p m_\s}{r_\b} J_{2,\ij} \left(\frac{R_\ij}{r_\b}\right)^2 P_2 (\vv{\hat r}_\b \cdot \vv{\hat \k}_\ij) \ , \llabel{130118y}
\end{eqnarray}
since terms in $m_\e / m_\ji (r_\b/r_\a)^3$ can also be neglected.

\section{Tidal potential}

\llabel{apenC}

The tidal potential of a body of mass $m_\ij$ when deformed by another body of mass $m'$ at the position $\vv{r}'$ is given by \citep[e.g.][]{Kaula_1964}:

\be 
V_{\ij} (\vv{r}, \vv{r}', m') = - k_{2,\ij} \frac{G m' R_\ij^5}{r^3 r'^3} P_2 (\vv{\hat r} \cdot \vv{\hat r}')
\llabel{130119a} \ ,
\ee
where we neglected terms in $(R_\ij/r)^4 (R_\ij/r')^4$.  
The resulting perturbing potential energy of a system composed of three bodies is given by:

\be
U_T = U_{T,\p} + U_{T,\s} + U_{T,\e}
\ , \llabel{130119b}
\ee
where we have for the planet
\be
U_{T,\e} = \sum_{\ij,\ji=0,1} m_\ij V_{\e} (\vv{r}_{\a \ij}, \vv{r}_{\a \ji}', m_\ji) = \sum_{\ij,\ji=0,1}  - k_{2,\e} \frac{G m_\ij m_\ji R_\e^5}{r_{\a \ij}^3 r_{\a \ji}'^3} P_2 (\vv{\hat r}_{\a \ij} \cdot \vv{\hat r}_{\a \ji}')
\ , \llabel{130119c}
\ee
and for each star ($\ij = 0,1$; $\jk=1-\ij$)
\begin{eqnarray}
U_{T,\ij} = 
m_\ji \left[ V_{\ij} (\vv{r}_\b, \vv{r}_\b', m_\ji) + V_{\ij} (\vv{r}_\b, \vv{r}_{\a \ij}', m_\e) \right]
+ m_\e  \left[ V_{\ij} (\vv{r}_{\a \ij}, \vv{r}_\b', m_\ji) + V_{\ij} (\vv{r}_{\a \ij}, \vv{r}_{\a \ij}', m_\e) \right]
\ . \llabel{130122a}
\end{eqnarray}
Neglecting the tidal interactions with the external body $m_\e$, i.e., neglecting terms in $m_\e / m_\ji (r_\b/r_\a)^3$, the above potential can be simplified as 
\be
U_{T,\ij} \approx m_\ij  V_{\ij} (\vv{r}_\b, \vv{r}_\b', m_\ji) = - k_{2,\ij} \frac{G m_\ji^2 R_\ij^5}{r_\b^3 r_\b'^3} P_2 (\vv{\hat r}_\b \cdot \vv{\hat r}_\b')
\ . \llabel{130122b}
\ee

Using expression (\ref{130118e}) we can rewrite
\begin{eqnarray}
\frac{P_2 (\vv{\hat r}_{\a \ij} \cdot \vv{\hat r}_{\a \ji}')}{r_{\a \ij}^3 r_{\a \ji}'^3} &\approx&
\frac{P_2 (\vv{\hat r}_\a \cdot \vv{\hat r}_\a')}{r_\a^3 r_\a'^3} 
+ \frac{3 \delta_\ij}{2 r_\a^3 r_\a'^3} \frac{r_\b}{r_\a} \left[ 
\vv{\hat r}_\b \cdot \vv{\hat r}_\a
- 5 (\vv{\hat r}_\b \cdot \vv{\hat r}_\a) (\vv{\hat r}_\a \cdot \vv{\hat r}_\a')^2
+ 2 (\vv{\hat r}_\b \cdot \vv{\hat r}_\a')(\vv{\hat r}_\a \cdot \vv{\hat r}_\a') \right]
\crm
&& + \frac{3 \delta_\ji}{2 r_\a^3 r_\a'^3} \frac{r_\b'}{r_\a'} \left[ 
\vv{\hat r}_\b' \cdot \vv{\hat r}_\a' 
- 5 (\vv{\hat r}_\b' \cdot \vv{\hat r}_\a')(\vv{\hat r}_\a \cdot \vv{\hat r}_\a')^2
+ 2 (\vv{\hat r}_\a \cdot \vv{\hat r}_\b')(\vv{\hat r}_\a \cdot \vv{\hat r}_\a')\right]
 \ , \llabel{130122c}
\end{eqnarray}
where we neglected terms in $(r_\b / r_\a)^2$, that is, we neglect terms in $(R_\e/r_\a)^6 (r_\b / r_\a)^2$ in the potential energy.
Replacing in expression (\ref{130119c}) we get for the planet
\be
U_{T,\e} =  - k_{2,\e} \frac{G \m^2 R_\e^5}{r_\a^3 r_\a'^3} P_2 (\vv{\hat r}_\a \cdot \vv{\hat r}_\a')
\ , \llabel{130122d}
\ee
since $m_\p^2 \delta_\p + m_\p m_\s (\delta_\p + \delta_\s) + m_\s^2 \delta_\s = 0$.

\section{Averaged quantities}

\llabel{apenA}

For completeness, we gather here the average formulae that are used in
the computation of secular equations. Let $F(\vv{r},\dot \vv{r})$ be
a function of a position vector $\vv{r}$ and velocity $\dot \vv{r}$, its
averaged expression over the mean anomaly $(M)$ is given by
\be
\moy{F}_{M} = \frac{1}{2\pi}\int_0^{2\pi} F(\vv{r}, \dot \vv{r})\, \d M\ .
\ee
Depending on the case, this integral is computing using the eccentric
anomaly $(E)$, or the true anomaly $(v)$ as an intermediate variable. The
basic formulae are
\begin{eqnarray}
&\d M = \Frac{r}{a}\d E = \Frac{r^2}{a^2\sqrt{1-e^2}}\d v\ , \crm
&\vv{r} = a(\cos E-e)\, \hat \vv{e} + a\sqrt{1-e^2}(\sin E)\, \hat \vv{k}
\times \hat \vv{e}\ , \crm
&\vv{r} = r\cos v\, \hat \vv{e} + r\sin v\, \hat \vv{k} \times \hat
\vv{e}\ , \crm
&\dot \vv{r} = \Frac{na}{\sqrt{1-e^2}}\, \hat \vv{k} \times 
(\hat \vv{r} + \vv{e})\ , \crm
&r = a (1-e\cos E) = \Frac{a(1-e^2)}{1+e\cos v}\ ,
\end{eqnarray}
where $\hat \vv{k}$ is the unit vector of the orbital angular momentum,
and $\vv{e}$ the Laplace-Runge-Lenz vector (Eq.\,\ref{100119a}). We have then
\be
\moy{\Frac{1}{r^3}} = \Frac{1}{a^3(1-e^2)^{3/2}}\ ,
\quad {\rm and } \quad
\moy{\frac{\vv{r} \trans{\vv{r}}}{r^5}} = 
\Frac{1}{2a^3(1-e^2)^{3/2}} \left(1-\hat \vv{k} \trans{\hat \vv{k}} \right) \ ,
\ee
where $\trans{\vv{u}}$ denotes the transpose of any vector $\vv{u}$. 
This leads to
\be
\moy{\Frac{1}{r^3}P_2(\hat \vv{r} \cdot \hat \vv{u})}
= -\Frac{1}{2a^3(1-e^2)^{3/2}} P_2(\hat \vv{k} \cdot \hat \vv{u})\ ,
\ee
for any unit vector $\hat \vv{u}$. In the same way,
\be
\moy{r^2} = a^2\left(1+\frac{3}{2}e^2\right)\ ,
\quad {\rm and } \quad
\moy{\vv{r} \trans{\vv{r}}} = a^2\Frac{1-e^2}{2}
\left(1-\hat \vv{k} \trans{\hat \vv{k}}\right) + \frac{5}{2} a^2 \vv{e}
\trans{\vv{e}}\ ,
\ee
give
\be
\moy{r^2 P_2(\hat \vv{r} \cdot \hat \vv{u})} = 
-\Frac{a^2}{2}\Big((1-e^2)P_2(\hat \vv{k} \cdot \hat \vv{u}) - 
5 e^2 P_2(\hat \vv{e} \cdot \hat \vv{u})\Big)\ .
\ee
The other useful formulae are
\be
\moy{\Frac{1}{r^6}} = \frac{1}{a^6} f_1(e)\ ,
\ee
\be
\moy{\Frac{1}{r^8}} = \frac{1}{a^8\sqrt{1-e^2}} f_2(e)\ ,
\ee
\be
\moy{\Frac{\vv{r} \trans{\vv{r}}}{r^8}} = 
\Frac{\sqrt{1-e^2}}{2a^6} f_4(e) \left(1-\hat \vv{k} \trans{\hat \vv{k}}\right)
+\frac{6+e^2}{4a^6(1-e^2)^{9/2}} \vv{e} \trans{\vv{e}}\ ,
\ee
\be
\moy{\Frac{\vv{r}}{r^8}} = \frac{5}{2}\Frac{1}{a^7\sqrt{1-e^2}}\, f_4(e)
\vv{e}\ ,
\ee
\be
\moy{\Frac{\vv{r}}{r^{10}}} = \frac{7}{2}\Frac{1}{a^9(1-e^2)}\, f_5(e)
\vv{e}\ ,
\ee
\be
\moy{\Frac{(\vv{r} \cdot \dot \vv{r}) \vv{r}}{r^{10}}} = \Frac{n}{2a^7
\sqrt{1-e^2}} f_5(e)\, \hat \vv{k} \times \vv{e}\ ,
\ee
where the $f_\ij (e)$ functions are given by expressions (\ref{090514n}) to
(\ref{090515e}).

Finally, for the average over the argument of the pericenter ($\w$), 
we can proceed in an identical manner:
\be
\moy{\vv{e} \trans{\vv{e}}}_\w = \frac{1}{2 \pi} \int_0^{2 \pi} \vv{e}
\trans{\vv{e}} \, d \w = \frac{e^2}{2} \left( 1 - \vv{k} \trans{\vv{k}}
\right) \ ,
\ee
which gives
\be
\moy{ \left( \vv{e} \cdot \hat \vv{u} \right) \vv{e}}_\w = \frac{e^2}{2}
\Big( \hat \vv{u} - ( \vv{k} \cdot \hat \vv{u} ) \vv{k} \Big) \ .
\ee

\bibliographystyle{spbasic}      
\bibliography{correia}   

\begin{thebibliography}{67}
\providecommand{\natexlab}[1]{#1}
\providecommand{\url}[1]{{#1}}
\providecommand{\urlprefix}{URL }
\expandafter\ifx\csname urlstyle\endcsname\relax
  \providecommand{\doi}[1]{DOI~\discretionary{}{}{}#1}\else
  \providecommand{\doi}{DOI~\discretionary{}{}{}\begingroup
  \urlstyle{rm}\Url}\fi
\providecommand{\eprint}[2][]{\url{#2}}

\bibitem[{{Alexander}(1973)}]{Alexander_1973}
{Alexander} ME (1973) {The Weak Friction Approximation and Tidal Evolution in
  Close Binary Systems}. \apss 23:459--510, \doi{10.1007/BF00645172}

\bibitem[{{Beaug{\'e}} and {Nesvorn{\'y}}(2012)}]{Beauge_Nesvorny_2012}
{Beaug{\'e}} C, {Nesvorn{\'y}} D (2012) {Multiple-planet Scattering and the
  Origin of Hot Jupiters}. \apj 751:119, \doi{10.1088/0004-637X/751/2/119},
  \eprint{1110.4392}

\bibitem[{{Bosanac} et~al(2015){Bosanac}, {Howell}, and
  {Fischbach}}]{Bosanac_etal_2015}
{Bosanac} N, {Howell} KC, {Fischbach} E (2015) {Stability of orbits near large
  mass ratio binary systems}. Celestial Mechanics and Dynamical Astronomy
  122:27--52, \doi{10.1007/s10569-015-9607-6}

\bibitem[{{Bou{\'e}} and {Fabrycky}(2014)}]{Boue_Fabrycky_2014b}
{Bou{\'e}} G, {Fabrycky} DC (2014) {Compact Planetary Systems Perturbed by an
  Inclined Companion. II. Stellar Spin-Orbit Evolution}. \apj 789:111,
  \doi{10.1088/0004-637X/789/2/111}, \eprint{1405.7636}

\bibitem[{{Bou{\'e}} and {Laskar}(2006)}]{Boue_Laskar_2006}
{Bou{\'e}} G, {Laskar} J (2006) {Precession of a planet with a satellite}.
  Icarus 185:312--330, \doi{10.1016/j.icarus.2006.07.019}

\bibitem[{{Bou{\'e}} and {Laskar}(2009)}]{Boue_Laskar_2009}
{Bou{\'e}} G, {Laskar} J (2009) {Spin axis evolution of two interacting
  bodies}. Icarus 201:750--767, \doi{10.1016/j.icarus.2009.02.001}

\bibitem[{{Brozovi{\'c}} et~al(2015){Brozovi{\'c}}, {Showalter}, {Jacobson},
  and {Buie}}]{Brozovic_etal_2015}
{Brozovi{\'c}} M, {Showalter} MR, {Jacobson} RA, {Buie} MW (2015) {The orbits
  and masses of satellites of Pluto}. \icarus 246:317--329,
  \doi{10.1016/j.icarus.2014.03.015}

\bibitem[{{Colombo}(1966)}]{Colombo_1966}
{Colombo} G (1966) {Cassini's second and third laws}. \aj 71:891--896

\bibitem[{{Correia}(2009)}]{Correia_2009}
{Correia} ACM (2009) {Secular Evolution of a Satellite by Tidal Effect:
  Application to Triton}. \apjl 704:L1--L4, \doi{10.1088/0004-637X/704/1/L1},
  \eprint{0909.4210}

\bibitem[{{Correia}(2015)}]{Correia_2015}
{Correia} ACM (2015) {Stellar and planetary Cassini states}. \aap 582:A69,
  \doi{10.1051/0004-6361/201525939}

\bibitem[{{Correia}(2016)}]{Correia_2016}
{Correia} ACM (2016) {Cassini states for black hole binaries}. \mnras
  457:L49--L53, \doi{10.1093/mnrasl/slv198}, \eprint{1511.01890}

\bibitem[{{Correia} and {Laskar}(2010)}]{Correia_Laskar_2010B}
{Correia} ACM, {Laskar} J (2010) {Tidal Evolution of Exoplanets}. In:
  Exoplanets, University of Arizona Press, pp 534--575

\bibitem[{{Correia} et~al(2011){Correia}, {Laskar}, {Farago}, and
  {Bou{\'e}}}]{Correia_etal_2011}
{Correia} ACM, {Laskar} J, {Farago} F, {Bou{\'e}} G (2011) {Tidal evolution of
  hierarchical and inclined systems}. Celestial Mechanics and Dynamical
  Astronomy 111:105--130, \doi{10.1007/s10569-011-9368-9}, \eprint{1107.0736}

\bibitem[{{Correia} et~al(2012){Correia}, {Bou{\'e}}, and
  {Laskar}}]{Correia_etal_2012}
{Correia} ACM, {Bou{\'e}} G, {Laskar} J (2012) {Pumping the Eccentricity of
  Exoplanets by Tidal Effect}. \apjl 744:L23,
  \doi{10.1088/2041-8205/744/2/L23}, \eprint{1111.5486}

\bibitem[{{Correia} et~al(2013){Correia}, {Bou{\'e}}, {Laskar}, and
  {Morais}}]{Correia_etal_2013}
{Correia} ACM, {Bou{\'e}} G, {Laskar} J, {Morais} MHM (2013) {Tidal damping of
  the mutual inclination in hierarchical systems}. \aap 553:A39,
  \doi{10.1051/0004-6361/201220482}, \eprint{1303.0864}

\bibitem[{{Correia} et~al(2014){Correia}, {Bou{\'e}}, {Laskar}, and
  {Rodr{\'{\i}}guez}}]{Correia_etal_2014}
{Correia} ACM, {Bou{\'e}} G, {Laskar} J, {Rodr{\'{\i}}guez} A (2014)
  {Deformation and tidal evolution of close-in planets and satellites using a
  Maxwell viscoelastic rheology}. \aap 571:A50,
  \doi{10.1051/0004-6361/201424211}, \eprint{1411.1860}

\bibitem[{{Correia} et~al(2015){Correia}, {Leleu}, {Rambaux}, and
  {Robutel}}]{Correia_etal_2015}
{Correia} ACM, {Leleu} A, {Rambaux} N, {Robutel} P (2015) {Spin-orbit coupling
  and chaotic rotation for circumbinary bodies. Application to the small
  satellites of the Pluto-Charon system}. \aap 580:L14,
  \doi{10.1051/0004-6361/201526800}, \eprint{1506.06733}

\bibitem[{{Couetdic} et~al(2010){Couetdic}, {Laskar}, {Correia}, {Mayor}, and
  {Udry}}]{Couetdic_etal_2010}
{Couetdic} J, {Laskar} J, {Correia} ACM, {Mayor} M, {Udry} S (2010) {Dynamical
  stability analysis of the HD 202206 system and constraints to the planetary
  orbits}. \aap 519:A10, \doi{10.1051/0004-6361/200913635}, \eprint{0911.1963}

\bibitem[{{Doolin} and {Blundell}(2011)}]{Doolin_Blundell_2011}
{Doolin} S, {Blundell} KM (2011) {The dynamics and stability of circumbinary
  orbits}. \mnras 418:2656--2668, \doi{10.1111/j.1365-2966.2011.19657.x},
  \eprint{1108.4144}

\bibitem[{{Doyle} et~al(2011){Doyle}, {Carter}, {Fabrycky}, {Slawson},
  {Howell}, {Winn}, {Orosz}, {Prsa}, {Welsh}, {Quinn}, {Latham}, {Torres},
  {Buchhave}, {Marcy}, {Fortney}, {Shporer}, {Ford}, {Lissauer}, {Ragozzine},
  {Rucker}, {Batalha}, {Jenkins}, {Borucki}, {Koch}, {Middour}, {Hall},
  {McCauliff}, {Fanelli}, {Quintana}, {Holman}, {Caldwell}, {Still},
  {Stefanik}, {Brown}, {Esquerdo}, {Tang}, {Furesz}, {Geary}, {Berlind},
  {Calkins}, {Short}, {Steffen}, {Sasselov}, {Dunham}, {Cochran}, {Boss},
  {Haas}, {Buzasi}, and {Fischer}}]{Doyle_etal_2011}
{Doyle} LR, {Carter} JA, {Fabrycky} DC, {Slawson} RW, {Howell} SB, {Winn} JN,
  {Orosz} JA, {Prsa} A, {Welsh} WF, {Quinn} SN, {Latham} D, {Torres} G,
  {Buchhave} LA, {Marcy} GW, {Fortney} JJ, {Shporer} A, {Ford} EB, {Lissauer}
  JJ, {Ragozzine} D, {Rucker} M, {Batalha} N, {Jenkins} JM, {Borucki} WJ,
  {Koch} D, {Middour} CK, {Hall} JR, {McCauliff} S, {Fanelli} MN, {Quintana}
  EV, {Holman} MJ, {Caldwell} DA, {Still} M, {Stefanik} RP, {Brown} WR,
  {Esquerdo} GA, {Tang} S, {Furesz} G, {Geary} JC, {Berlind} P, {Calkins} ML,
  {Short} DR, {Steffen} JH, {Sasselov} D, {Dunham} EW, {Cochran} WD, {Boss} A,
  {Haas} MR, {Buzasi} D, {Fischer} D (2011) {Kepler-16: A Transiting
  Circumbinary Planet}. Science 333:1602--, \doi{10.1126/science.1210923},
  \eprint{1109.3432}

\bibitem[{{Efroimsky} and {Williams}(2009)}]{Efroimsky_Williams_2009}
{Efroimsky} M, {Williams} JG (2009) {Tidal torques: a critical review of some
  techniques}. Celestial Mechanics and Dynamical Astronomy 104:257--289,
  \doi{10.1007/s10569-009-9204-7}, \eprint{0803.3299}

\bibitem[{{Eggleton} and {Kiseleva-Eggleton}(2001)}]{Eggleton_Kiseleva_2001}
{Eggleton} PP, {Kiseleva-Eggleton} L (2001) {Orbital Evolution in Binary and
  Triple Stars, with an Application to SS Lacertae}. \apj 562:1012--1030,
  \doi{10.1086/323843}, \eprint{arXiv:astro-ph/0104126}

\bibitem[{{Farago} and {Laskar}(2010)}]{Farago_Laskar_2010}
{Farago} F, {Laskar} J (2010) {High-inclination orbits in the secular
  quadrupolar three-body problem}. \mnras 401:1189--1198,
  \doi{10.1111/j.1365-2966.2009.15711.x}, \eprint{0909.2287}

\bibitem[{{Ferraz-Mello}(2013)}]{Ferraz-Mello_2013}
{Ferraz-Mello} S (2013) {Tidal synchronization of close-in satellites and
  exoplanets. A rheophysical approach}. Celestial Mechanics and Dynamical
  Astronomy 116:109--140, \doi{10.1007/s10569-013-9482-y}, \eprint{1204.3957}

\bibitem[{{Ford} et~al(2000{\natexlab{a}}){Ford}, {Joshi}, {Rasio}, and
  {Zbarsky}}]{Ford_etal_2000p}
{Ford} EB, {Joshi} KJ, {Rasio} FA, {Zbarsky} B (2000{\natexlab{a}})
  {Theoretical Implications of the PSR B1620-26 Triple System and Its Planet}.
  \apj 528:336--350, \doi{10.1086/308167}, \eprint{astro-ph/9905347}

\bibitem[{{Ford} et~al(2000{\natexlab{b}}){Ford}, {Kozinsky}, and
  {Rasio}}]{Ford_etal_2000}
{Ford} EB, {Kozinsky} B, {Rasio} FA (2000{\natexlab{b}}) {Secular Evolution of
  Hierarchical Triple Star Systems}. \apj 535:385--401, \doi{10.1086/308815}

\bibitem[{{Goldreich}(1966)}]{Goldreich_1966a}
{Goldreich} P (1966) {History of the Lunar Orbit}. Reviews of Geophysics and
  Space Physics 4:411--439, \doi{10.1029/RG004i004p00411}

\bibitem[{{Goldstein}(1950)}]{Goldstein_1950}
{Goldstein} H (1950) {Classical mechanics}. Addison-Wesley, Reading

\bibitem[{{Harrington}(1968)}]{Harrington_1968}
{Harrington} RS (1968) {Dynamical evolution of triple stars.} \aj 73:190--194,
  \doi{10.1086/110614}

\bibitem[{{Hut}(1980)}]{Hut_1980}
{Hut} P (1980) {Stability of tidal equilibrium}. \aap 92:167--170

\bibitem[{{Kaula}(1964)}]{Kaula_1964}
{Kaula} WM (1964) {Tidal dissipation by solid friction and the resulting
  orbital evolution}. \rg 2:661--685

\bibitem[{{Kennedy} et~al(2012){Kennedy}, {Wyatt}, {Sibthorpe}, {Duch{\^e}ne},
  {Kalas}, {Matthews}, {Greaves}, {Su}, and {Fitzgerald}}]{Kennedy_etal_2012a}
{Kennedy} GM, {Wyatt} MC, {Sibthorpe} B, {Duch{\^e}ne} G, {Kalas} P, {Matthews}
  BC, {Greaves} JS, {Su} KYL, {Fitzgerald} MP (2012) {99 Herculis: host to a
  circumbinary polar-ring debris disc}. \mnras 421:2264--2276,
  \doi{10.1111/j.1365-2966.2012.20448.x}, \eprint{1201.1911}

\bibitem[{{Kidder}(1995)}]{Kidder_1995}
{Kidder} LE (1995) {Coalescing binary systems of compact objects to
  (post)$^{5/2}$-Newtonian order. V. Spin effects}. \prd 52:821--847,
  \doi{10.1103/PhysRevD.52.821}, \eprint{gr-qc/9506022}

\bibitem[{{Kostov} et~al(2015){Kostov}, {Orosz}, {Welsh}, {Doyle}, {Fabrycky},
  {Haghighipour}, {Quarles}, {Short}, {Cochran}, {Endl}, {Ford}, {Gregorio},
  {Hinse}, {Isaacson}, {Jenkins}, {Jensen}, {Kull}, {Latham}, {Lissauer},
  {Marcy}, {Mazeh}, {Muller}, {Pepper}, {Quinn}, {Ragozzine}, {Shporer},
  {Steffen}, {Torres}, {Windmiller}, and {Borucki}}]{Kostov_etal_2016}
{Kostov} VB, {Orosz} JA, {Welsh} WF, {Doyle} LR, {Fabrycky} DC, {Haghighipour}
  N, {Quarles} B, {Short} DR, {Cochran} WD, {Endl} M, {Ford} EB, {Gregorio} J,
  {Hinse} TC, {Isaacson} H, {Jenkins} JM, {Jensen} ELN, {Kull} I, {Latham} DW,
  {Lissauer} JJ, {Marcy} GW, {Mazeh} T, {Muller} TWA, {Pepper} J, {Quinn} SN,
  {Ragozzine} D, {Shporer} A, {Steffen} JH, {Torres} G, {Windmiller} G,
  {Borucki} WJ (2015) {KOI-2939b: the largest and longest-period Kepler
  transiting circumbinary planet}. ArXiv e-prints \eprint{1512.00189}

\bibitem[{{Kozai}(1962)}]{Kozai_1962}
{Kozai} Y (1962) {Secular perturbations of asteroids with high inclination and
  eccentricity}. \aj 67:591--598, \doi{10.1086/108790}

\bibitem[{{Lainey} et~al(2009){Lainey}, {Arlot}, {Karatekin}, and {van
  Hoolst}}]{Lainey_etal_2009}
{Lainey} V, {Arlot} JE, {Karatekin} {\"O}, {van Hoolst} T (2009) {Strong tidal
  dissipation in Io and Jupiter from astrometric observations}. Nature
  459:957--959, \doi{10.1038/nature08108}

\bibitem[{{Lainey} et~al(2012){Lainey}, {Karatekin}, {Desmars}, {Charnoz},
  {Arlot}, {Emelyanov}, {Le Poncin-Lafitte}, {Mathis}, {Remus}, {Tobie}, and
  {Zahn}}]{Lainey_etal_2012}
{Lainey} V, {Karatekin} {\"O}, {Desmars} J, {Charnoz} S, {Arlot} JE,
  {Emelyanov} N, {Le Poncin-Lafitte} C, {Mathis} S, {Remus} F, {Tobie} G,
  {Zahn} JP (2012) {Strong Tidal Dissipation in Saturn and Constraints on
  Enceladus' Thermal State from Astrometry}. \apj 752:14,
  \doi{10.1088/0004-637X/752/1/14}, \eprint{1204.0895}

\bibitem[{{Lambeck}(1988)}]{Lambeck_1988}
{Lambeck} K (1988) {Geophysical geodesy : the slow deformations of the earth
  Lambeck.} Oxford [England] : Clarendon Press ; New York : Oxford University
  Press, 1988.

\bibitem[{{Laskar}(2000)}]{Laskar_2000}
{Laskar} J (2000) {On the Spacing of Planetary Systems}. Physical Review
  Letters 84:3240--3243

\bibitem[{{Laskar} and {Bou{\'e}}(2010)}]{Laskar_Boue_2010}
{Laskar} J, {Bou{\'e}} G (2010) {Explicit expansion of the three-body
  disturbing function for arbitrary eccentricities and inclinations}. \aap
  522:A60, \doi{10.1051/0004-6361/201014496}, \eprint{1008.2947}

\bibitem[{{Laskar} et~al(2004){Laskar}, {Robutel}, {Joutel}, {Gastineau},
  {Correia}, and {Levrard}}]{Laskar_etal_2004E}
{Laskar} J, {Robutel} P, {Joutel} F, {Gastineau} M, {Correia} ACM, {Levrard} B
  (2004) {A long-term numerical solution for the insolation quantities of the
  Earth}. \aap 428:261--285, \doi{10.1051/0004-6361:20041335}

\bibitem[{{Lee} and {Peale}(2003)}]{Lee_Peale_2003}
{Lee} MH, {Peale} SJ (2003) {Secular Evolution of Hierarchical Planetary
  Systems}. \apj 592:1201--1216, \doi{10.1086/375857},
  \eprint{arXiv:astro-ph/0304454}

\bibitem[{{Lidov}(1962)}]{Lidov_1962}
{Lidov} ML (1962) {The evolution of orbits of artificial satellites of planets
  under the action of gravitational perturbations of external bodies}. \planss
  9:719--759, \doi{10.1016/0032-0633(62)90129-0}

\bibitem[{{Lidov} and {Ziglin}(1976)}]{Lidov_Ziglin_1976}
{Lidov} ML, {Ziglin} SL (1976) {Non-restricted double-averaged three body
  problem in Hill's case}. Celestial Mechanics 13:471--489,
  \doi{10.1007/BF01229100}

\bibitem[{{MacDonald}(1964)}]{MacDonald_1964}
{MacDonald} GJF (1964) {Tidal friction}. Revs Geophys 2:467--541

\bibitem[{{Makarov}(2015)}]{Makarov_2015}
{Makarov} VV (2015) {Equilibrium rotation of semiliquid exoplanets and
  satellites}. ArXiv e-prints \eprint{1507.07383}

\bibitem[{{Marchal}(1990)}]{Marchal_1990}
{Marchal} C (1990) {The Three-Body Problem}. Elsevier, Amsterdam

\bibitem[{{Martin} and {Triaud}(2014)}]{Martin_Triaud_2014}
{Martin} DV, {Triaud} AHMJ (2014) {Planets transiting non-eclipsing binaries}.
  \aap 570:A91, \doi{10.1051/0004-6361/201323112}, \eprint{1404.5360}

\bibitem[{{Migaszewski}(2012)}]{Migaszewski_2012}
{Migaszewski} C (2012) {The generalized non-conservative model of a 1-planet
  system revisited}. Celestial Mechanics and Dynamical Astronomy 113:169--203,
  \doi{10.1007/s10569-012-9413-3}, \eprint{1203.2358}

\bibitem[{{Migaszewski} and
  {Go{\'z}dziewski}(2011)}]{Migaszewski_Gozdziewski_2011}
{Migaszewski} C, {Go{\'z}dziewski} K (2011) {The non-resonant, relativistic
  dynamics of circumbinary planets}. \mnras 411:565--583,
  \doi{10.1111/j.1365-2966.2010.17702.x}, \eprint{1006.5961}

\bibitem[{{Mignard}(1979)}]{Mignard_1979}
{Mignard} F (1979) {The evolution of the lunar orbit revisited. I}. Moon and
  Planets 20:301--315

\bibitem[{{Naoz} et~al(2011){Naoz}, {Farr}, {Lithwick}, {Rasio}, and
  {Teyssandier}}]{Naoz_etal_2011}
{Naoz} S, {Farr} WM, {Lithwick} Y, {Rasio} FA, {Teyssandier} J (2011) {Hot
  Jupiters from secular planet-planet interactions}. \nat 473:187--189,
  \doi{10.1038/nature10076}, \eprint{1011.2501}

\bibitem[{{Orosz} et~al(2012{\natexlab{a}}){Orosz}, {Welsh}, {Carter},
  {Brugamyer}, {Buchhave}, {Cochran}, {Endl}, {Ford}, {MacQueen}, {Short},
  {Torres}, {Windmiller}, {Agol}, {Barclay}, {Caldwell}, {Clarke}, {Doyle},
  {Fabrycky}, {Geary}, {Haghighipour}, {Holman}, {Ibrahim}, {Jenkins},
  {Kinemuchi}, {Li}, {Lissauer}, {Pr{\v s}a}, {Ragozzine}, {Shporer}, {Still},
  and {Wade}}]{Orosz_etal_2012a}
{Orosz} JA, {Welsh} WF, {Carter} JA, {Brugamyer} E, {Buchhave} LA, {Cochran}
  WD, {Endl} M, {Ford} EB, {MacQueen} P, {Short} DR, {Torres} G, {Windmiller}
  G, {Agol} E, {Barclay} T, {Caldwell} DA, {Clarke} BD, {Doyle} LR, {Fabrycky}
  DC, {Geary} JC, {Haghighipour} N, {Holman} MJ, {Ibrahim} KA, {Jenkins} JM,
  {Kinemuchi} K, {Li} J, {Lissauer} JJ, {Pr{\v s}a} A, {Ragozzine} D, {Shporer}
  A, {Still} M, {Wade} RA (2012{\natexlab{a}}) {The Neptune-sized Circumbinary
  Planet Kepler-38b}. \apj 758:87, \doi{10.1088/0004-637X/758/2/87},
  \eprint{1208.3712}

\bibitem[{{Orosz} et~al(2012{\natexlab{b}}){Orosz}, {Welsh}, {Carter},
  {Fabrycky}, {Cochran}, {Endl}, {Ford}, {Haghighipour}, {MacQueen}, {Mazeh},
  {Sanchis-Ojeda}, {Short}, {Torres}, {Agol}, {Buchhave}, {Doyle}, {Isaacson},
  {Lissauer}, {Marcy}, {Shporer}, {Windmiller}, {Barclay}, {Boss}, {Clarke},
  {Fortney}, {Geary}, {Holman}, {Huber}, {Jenkins}, {Kinemuchi}, {Kruse},
  {Ragozzine}, {Sasselov}, {Still}, {Tenenbaum}, {Uddin}, {Winn}, {Koch}, and
  {Borucki}}]{Orosz_etal_2012b}
{Orosz} JA, {Welsh} WF, {Carter} JA, {Fabrycky} DC, {Cochran} WD, {Endl} M,
  {Ford} EB, {Haghighipour} N, {MacQueen} PJ, {Mazeh} T, {Sanchis-Ojeda} R,
  {Short} DR, {Torres} G, {Agol} E, {Buchhave} LA, {Doyle} LR, {Isaacson} H,
  {Lissauer} JJ, {Marcy} GW, {Shporer} A, {Windmiller} G, {Barclay} T, {Boss}
  AP, {Clarke} BD, {Fortney} J, {Geary} JC, {Holman} MJ, {Huber} D, {Jenkins}
  JM, {Kinemuchi} K, {Kruse} E, {Ragozzine} D, {Sasselov} D, {Still} M,
  {Tenenbaum} P, {Uddin} K, {Winn} JN, {Koch} DG, {Borucki} WJ
  (2012{\natexlab{b}}) {Kepler-47: A Transiting Circumbinary Multiplanet
  System}. Science 337:1511--, \doi{10.1126/science.1228380},
  \eprint{1208.5489}

\bibitem[{{Penev} et~al(2012){Penev}, {Jackson}, {Spada}, and
  {Thom}}]{Penev_etal_2012}
{Penev} K, {Jackson} B, {Spada} F, {Thom} N (2012) {Constraining Tidal
  Dissipation in Stars from the Destruction Rates of Exoplanets}. \apj 751:96,
  \doi{10.1088/0004-637X/751/2/96}, \eprint{1205.1803}

\bibitem[{{Plavchan} et~al(2013){Plavchan}, {G{\"u}th}, {Laohakunakorn}, and
  {Parks}}]{Plavchan_etal_2013}
{Plavchan} P, {G{\"u}th} T, {Laohakunakorn} N, {Parks} JR (2013) {The
  identification of 93 day periodic photometric variability for YSO YLW 16A}.
  \aap 554:A110, \doi{10.1051/0004-6361/201220747}, \eprint{1304.2398}

\bibitem[{Poincar{\'e}(1905)}]{Poincare_1905}
Poincar{\'e} H (1905) Le{\c c}ons de M{\'e}canique C{\'e}leste, Tome I.
  {Gauthier-Villars.} Paris

\bibitem[{{Singer}(1968)}]{Singer_1968}
{Singer} SF (1968) {The Origin of the Moon and Geophysical Consequences}.
  \gjras 15:205--226

\bibitem[{{Skumanich}(1972)}]{Skumanich_1972}
{Skumanich} A (1972) {Time Scales for CA II Emission Decay, Rotational Braking,
  and Lithium Depletion}. \apj 171:565, \doi{10.1086/151310}

\bibitem[{{Smart}(1953)}]{Smart_1953}
{Smart} WM (1953) {Celestial Mechanics.} London, New York, Longmans, Green

\bibitem[{{Touma} et~al(2009){Touma}, {Tremaine}, and
  {Kazandjian}}]{Touma_etal_2009}
{Touma} JR, {Tremaine} S, {Kazandjian} MV (2009) {Gauss's method for secular
  dynamics, softened}. \mnras 394:1085--1108,
  \doi{10.1111/j.1365-2966.2009.14409.x}, \eprint{0811.2812}

\bibitem[{{Verrier} and {Evans}(2009)}]{Verrier_Evans_2009}
{Verrier} PE, {Evans} NW (2009) {High-inclination planets and asteroids in
  multistellar systems}. \mnras 394:1721--1726,
  \doi{10.1111/j.1365-2966.2009.14446.x}, \eprint{0812.4528}

\bibitem[{{Ward}(1975)}]{Ward_1975}
{Ward} WR (1975) {Tidal friction and generalized Cassini's laws in the solar
  system}. \aj 80:64--70

\bibitem[{{Ward} and {Hamilton}(2004)}]{Ward_Hamilton_2004}
{Ward} WR, {Hamilton} DP (2004) {Tilting Saturn. I. Analytic Model}. \aj
  128:2501--2509, \doi{10.1086/424533}

\bibitem[{{Welsh} et~al(2012){Welsh}, {Orosz}, {Carter}, {Fabrycky}, {Ford},
  {Lissauer}, {Pr{\v s}a}, {Quinn}, {Ragozzine}, {Short}, {Torres}, {Winn},
  {Doyle}, {Barclay}, {Batalha}, {Bloemen}, {Brugamyer}, {Buchhave},
  {Caldwell}, {Caldwell}, {Christiansen}, {Ciardi}, {Cochran}, {Endl},
  {Fortney}, {Gautier}, {Gilliland}, {Haas}, {Hall}, {Holman}, {Howard},
  {Howell}, {Isaacson}, {Jenkins}, {Klaus}, {Latham}, {Li}, {Marcy}, {Mazeh},
  {Quintana}, {Robertson}, {Shporer}, {Steffen}, {Windmiller}, {Koch}, and
  {Borucki}}]{Welsh_etal_2012}
{Welsh} WF, {Orosz} JA, {Carter} JA, {Fabrycky} DC, {Ford} EB, {Lissauer} JJ,
  {Pr{\v s}a} A, {Quinn} SN, {Ragozzine} D, {Short} DR, {Torres} G, {Winn} JN,
  {Doyle} LR, {Barclay} T, {Batalha} N, {Bloemen} S, {Brugamyer} E, {Buchhave}
  LA, {Caldwell} C, {Caldwell} DA, {Christiansen} JL, {Ciardi} DR, {Cochran}
  WD, {Endl} M, {Fortney} JJ, {Gautier} TN III, {Gilliland} RL, {Haas} MR,
  {Hall} JR, {Holman} MJ, {Howard} AW, {Howell} SB, {Isaacson} H, {Jenkins} JM,
  {Klaus} TC, {Latham} DW, {Li} J, {Marcy} GW, {Mazeh} T, {Quintana} EV,
  {Robertson} P, {Shporer} A, {Steffen} JH, {Windmiller} G, {Koch} DG,
  {Borucki} WJ (2012) {Transiting circumbinary planets Kepler-34 b and
  Kepler-35 b}. \nat 481:475--479, \doi{10.1038/nature10768},
  \eprint{1204.3955}

\bibitem[{{Winn} et~al(2011){Winn}, {Albrecht}, {Johnson}, {Torres}, {Cochran},
  {Marcy}, {Howard}, {Isaacson}, {Fischer}, {Doyle}, {Welsh}, {Carter},
  {Fabrycky}, {Ragozzine}, {Quinn}, {Shporer}, {Howell}, {Latham}, {Orosz},
  {Prsa}, {Slawson}, {Borucki}, {Koch}, {Barclay}, {Boss},
  {Christensen-Dalsgaard}, {Girouard}, {Jenkins}, {Klaus}, {Meibom}, {Morris},
  {Sasselov}, {Still}, and {Van Cleve}}]{Winn_etal_2011}
{Winn} JN, {Albrecht} S, {Johnson} JA, {Torres} G, {Cochran} WD, {Marcy} GW,
  {Howard} AW, {Isaacson} H, {Fischer} D, {Doyle} L, {Welsh} W, {Carter} JA,
  {Fabrycky} DC, {Ragozzine} D, {Quinn} SN, {Shporer} A, {Howell} SB, {Latham}
  DW, {Orosz} J, {Prsa} A, {Slawson} RW, {Borucki} WJ, {Koch} D, {Barclay} T,
  {Boss} AP, {Christensen-Dalsgaard} J, {Girouard} FR, {Jenkins} J, {Klaus} TC,
  {Meibom} S, {Morris} RL, {Sasselov} D, {Still} M, {Van Cleve} J (2011)
  {Spin-Orbit Alignment for the Circumbinary Planet Host Kepler-16 A}. \apjl
  741:L1, \doi{10.1088/2041-8205/741/1/L1}, \eprint{1109.3198}

\bibitem[{{Yoder}(1995)}]{Yoder_1995cnt}
{Yoder} CF (1995) {Astrometric and geodetic properties of Earth and the Solar
  System}. In: Global Earth Physics: A Handbook of Physical Constants, American
  Geophysical Union, Washington D.C, pp 1--31

\end{thebibliography}

\end{document}